\documentclass[]{aa} 
\usepackage{txfonts}
\usepackage{natbib}
\usepackage{graphicx, xspace}
\usepackage{placeins}
\usepackage{array}
\usepackage[switch]{lineno}
\usepackage{hyperref}
\usepackage{color}
\usepackage{gensymb}

\makeatletter

\newcommand{\Rmnum}[1]{\expandafter\@slowromancap\romannumeral #1@}
\makeatother

\def\d{\,d$^{-1}$\xspace}
\begin{document}


\title{Forward seismic modelling of the pulsating magnetic B-type star HD\,43317}

\authorrunning{B.\,Buysschaert et al.}
\titlerunning{Asteroseismic modelling of HD\,43317}

\author{B.\,Buysschaert\inst{1,2}, 
C.\,Aerts\inst{2,3},
D.\,M.\,Bowman\inst{2},
C.\,Johnston\inst{2},
T.\,Van\,Reeth\inst{2},
M.\,G.\,Pedersen\inst{2}, \\
S.\,Mathis\inst{4,1}, and 
C.\,Neiner\inst{1}
}
\mail{bram.buysschaert@obspm.fr} 

\institute{LESIA, Observatoire de Paris, PSL Research University, CNRS, Sorbonne Universit\'es, UPMC Univ. Paris 06, Univ. Paris Diderot, Sorbonne Paris Cit\'e, 5 place Jules Janssen, F-92195 Meudon, France 
  \and Instituut voor Sterrenkunde, KU Leuven, Celestijnenlaan 200D, 3001 Leuven, Belgium 
  \and Dept. of Astrophysics, IMAPP, Radboud University Nijmegen, 6500 GL, Nijmegen, The Netherlands 
  \and Laboratoire AIM Paris-Saclay, CEA/DRF – CNRS – Universite Paris Diderot, IRFU/DAp Centre de Saclay, F-91191 Gif-sur-Yvette, France 
} 

\abstract{The large-scale magnetic fields detected at the surface of about 10\,\% of hot stars extend into the stellar interior, where they may alter the structure.  Deep inner regions of stars are only observable using asteroseismology.  Here, we investigated the pulsating magnetic B3.5V star HD\,43317, inferred its interior properties and assessed whether the dipolar magnetic field with a surface strength of $B_p = 1312 \pm 332$\,G caused different properties compared to those of non-magnetic stars.  We analysed the latest version of the star's 150\,d CoRoT light curve and extracted 35 significant frequencies, 28 of which were determined to be independent and not related to the known surface rotation period of $P_{\rm rot} = 0.897673$\,d.  We performed forward seismic modelling based on non-magnetic, non-rotating 1D MESA models and the adiabatic module of the pulsation code GYRE, utilizing a grid-based approach.  Our aim was to estimate the stellar mass, age, and convective core overshooting.  The GYRE calculations were done for uniform rotation with $P_{\rm rot}$.  This modelling was able to explain 16 of the 28 frequencies as gravity modes belonging to  retrograde modes with $(\ell, m) = (1, -1)$ and $(2, -1)$ period spacing patterns and one distinct prograde $(2,2)$ mode.  The modelling resulted in a stellar mass $M_{\star} = 5.8^{+0.1}_{-0.2}$\,$\mathrm{M_{\odot}}$, a central hydrogen mass fraction $X_c = 0.54^{+0.01}_{-0.02}$, and exponential convective core overshooting parameter $f_{\rm ov} = 0.004^{+0.014}_{-0.002}$.  The low value for $f_{\rm ov}$ is compatible with the suppression of near-core mixing due to a magnetic field but the uncertainties are too large to pinpoint such suppression as the sole physical interpretation.  We assessed the frequency shifts of pulsation modes caused by the Lorentz and the Coriolis forces and found magnetism to have a lower impact than rotation for this star. Including magnetism in future pulsation computations would be highly relevant to exploit current and future photometric time series spanning at least one year, such as those assembled by the {\it Kepler\/} space telescope and expected from the TESS (Continuous Viewing Zone) and PLATO space missions.}


\maketitle

\section{Introduction}
\label{sec:intro}
Large-scale magnetic fields are detected at the surface of about 10\,\% of the observed early-type stars, i.e., having spectral type O, B, or A \citep[see e.g.,][]{1968PASP...80..281W, 2007MsT..........1P, 2012ApJ...750....2S, 2014IAUS..302..265W, 2015IAUS..305...53G, 2018MNRAS.475.5144S}.  For the O, B and A stars (which we refer to as hot stars), most magnetic   detections are obtained by dedicated and systematic surveys using high-resolution spectropolarimetry \citep[e.g., MiMeS,][]{2016MNRAS.456....2W}. Magnetic candidates are discovered by indirect evidence of a large-scale magnetic field, such as rotational modulation caused by surface abundance inhomogeneities or magnetically confined matter in the circumstellar environment or by observed peculiar photospheric chemical abundances.

The majority of the detected large-scale magnetic fields have a simple geometry, often a magnetic dipole inclined to the rotation axis, with magnetic field strengths ranging from 100\,G up to a few tens of kG.  Because these large-scale magnetic fields are found to be stable over a time span of several decades and because their properties do not scale with stellar parameters, they are thought to be created during the star formation process, i.e., have a fossil origin \citep[e.g.,][]{1999stma.book.....M, 2015IAUS..305...61N}.  Results of both semi-analytical descriptions and numerical simulations demonstrated that the magnetic field detected at the surface must extend deep within the radiative envelope of hot stars to ensure the stability of these magnetic fields \citep[see e.g.,][]{1973MNRAS.163...77M, 2010ApJ...724L..34D, 2004Natur.431..819B, 2006A+A...450.1077B, 2007A+A...469..275B, 2008MNRAS.386.1947B, 2009ApJ...705.1000F, 2010A+A...517A..58D}.  These (large-scale) magnetic fields may influence the stellar structure due to the Lorentz force, in addition to the pressure force and gravity.  Theoretical work suggests that the internal properties of a large-scale magnetic field would impose uniform rotation within the radiative layers \citep[e.g.,][]{1937MNRAS..97..458F, 1992MNRAS.257..593M, 1999A+A...349..189S, 2005A+A...440..653M, 2011IAUS..272...14Z}.  As such, it is expected that matter loses its inertia quicker when overshooting the convective core boundary than in the absence of a magnetic field \citep[e.g.,][]{1981ApJ...245..286P, 2004ApJ...601..512B}.  Thus, the presence of magnetic forces should result in a smaller convective core overshooting region.

Aside from isochrone fitting of binaries or clusters,  asteroseismology is an excellent method to probe the internal stellar structure from the interpretation of detected stellar oscillations \citep{2010aste.book.....A}. This is particularly the case to estimate  core overshooting from gravity (g) modes, as these are most sensitive to the deep stellar interior.  About 15 magnetic pulsating stars with a spectral type ealier than B3 have been discovered so far \citep[e.g.,][]{2017IAUS..329..146B}. This is mainly due to the fact that only ${\sim}10\,\%$ of hot stars are magnetic and that seismic data are available for only a few such stars. In addition, the presence of a very strong magnetic field may influence the driving and damping of pulsation modes and waves \citep[e.g.,][]{2017MNRAS.466.2181L}.  Of the hottest known magnetic pulsating stars, only $\beta$\,Cep \citep{2000ApJ...531L.143S, 2013A+A...555A..46H} and V2052\,Oph \citep{2012A+A...537A.148N, 2012MNRAS.424.2380H, 2012MNRAS.427..483B} have been studied using combined asteroseismic and magnetometric analyses (i.e., magneto-asteroseismology).  These two stars are both pressure-mode (p-mode) pulsators.  The study of V2052\,Oph demonstrated that this magnetic pulsating star has a smaller convective core overshooting layer compared to the similar yet non-magnetic pulsator $\theta\,$Oph \citep{2007MNRAS.381.1482B}, as expected from theoretical predictions.

In this paper, we selected the B3.5V star HD\,43317 for detailed forward seismic modelling, because it has a well characterized large-scale magnetic field \citep{2013A+A...557L..16B, 2017A+A...605A.104B}, and, most importantly, its CoRoT light curve \citep[Convection Rotation and planetary Transits;][]{2006cosp...36.3749B} shows a rich frequency spectrum of many g-mode frequencies \citep{2012A+A...542A..55P}.  The spectroscopy of HD\,43317 suggested it to be a single star and the spectroscopic analysis indicated a solar-like metallicity, an effective temperature $T_{\rm eff} = 17350 \pm 750$\,K and a surface gravity $\log g = 4.0 \pm 0.1$\,dex \citep{2012A+A...542A..55P}.  Furthermore, the measured rotation velocity is $v \sin i = 115 \pm 9$\,km\,s$^{-1}$ \citep{2012A+A...542A..55P} and the rotation period is $P_{\rm rot} = 0.897673(4)$\,d \citep{2017A+A...605A.104B}.  From this work, it was also found that the dipolar magnetic field at the surface of HD\,43317 has a strength of $1.0 - 1.5$\,kG, which should be strong enough to result in uniform rotation in the radiative layer according to theoretical criteria \citep{2017A+A...605A.104B}.  We note that most studied B- and F-type g-mode pulsators with asteroseismic core-to-envelope rotation rates rotate nearly uniformly. The studied sample with such measurements contains stars that rotate up to about half their critical rotation rate but magnetic measurements are not available for them (\citealt{2017ApJ...847L...7A} and Van Reeth et al.\ submitted).

In this paper, we assess the interior properties of HD\,43317 from its g modes. Since the CoRoT data underwent a final end-of-life reduction to produce the version of the light curves in the public data archive, we performed a new frequency extraction in Sect.\,\ref{sec:frequency} to obtain a conservative list of candidate pulsation mode frequencies.  Section\,\ref{sec:modelling} treats the detailed forward modelling of HD\,43317 by coupling 1D stellar structure models with pulsation computations, adopting various hypotheses as explained in the text.  We discuss the results and implications of the forward modelling in Sect.\,\ref{sec:discussion}, and summarize and draw conclusions in Sect.\,\ref{sec:conclusion}.

\section{Frequency extraction}
\label{sec:frequency}
\subsection{CoRoT light curve}
CoRoT observed HD\,43317 during campaign LRa03 from 30/09/2009 until 01/03/2010, producing high-cadence, high-precision space-based photometry \citep[see ][for explicit details on the spacecraft and its performance]{2009A+A...506..411A}.  The total light curve spanned 150.49\,d with a median observing cadence of 32\,s and was retrieved from the CoRoT public archive\footnote{Available at \url{http://idoc-corot.ias.u-psud.fr/sitools/client-user/COROT_N2_PUBLIC_DATA/project-index.html}.}. An earlier version of the light curve was analysed by \citet{2012A+A...542A..55P} to study the photometric variability of HD\,43317.

\begin{figure*}[t]
\centering
\includegraphics[width=\textwidth, height = 0.50\textheight]{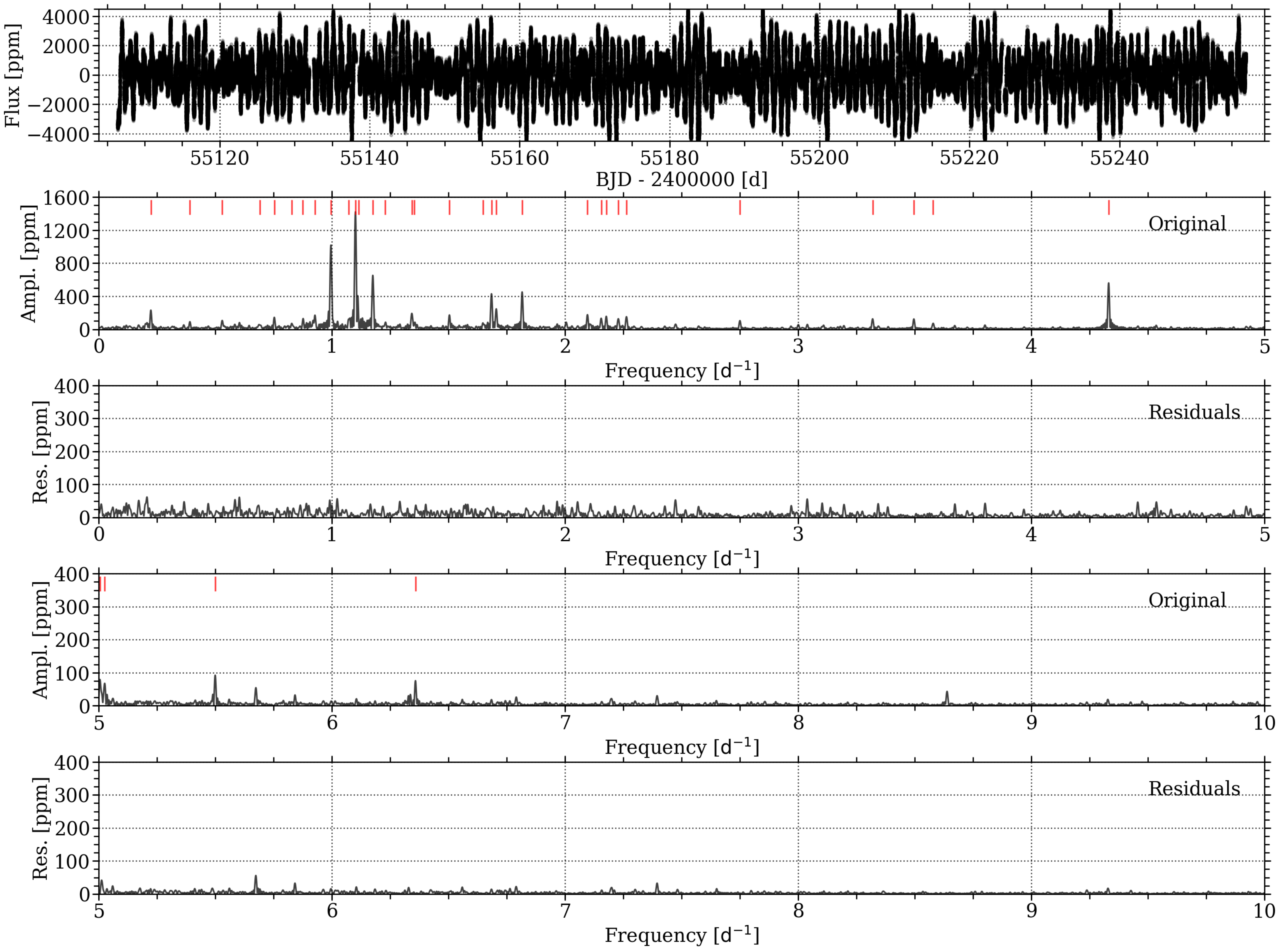}%
\caption{CoRoT light curve (\textit{top}) and the corresponding Lomb-Scargle periodograms (\textit{2nd} and \textit{4th row} for increased visibility) of HD\,43317.  The 35 significant frequencies are indicated by the red tick marks in the upper parts of the periodogram panels.  Corresponding periodograms of the residual light curve are given in panels on the \textit{3rd} and \textit{5th row}.}
\label{fig:CoRoT}
\end{figure*}

We re-processed the CoRoT light curve by removing obvious outliers indicated by non-zero status flag. We also removed the so-called ``in-pasted'' observations added by the CoRoT data centre to fill observing gaps. Further, we performed a correction for the (likely instrumental) long-term variability through application of a local linear regression filter, removing all signal with a frequency below 0.05\,\d.  We ensured that the periodic variability detected in the data was well above this value and that the amplitudes of the frequencies were not significantly altered by the applied correction. Finally, we converted the flux to parts-per-million (ppm).  This corrected light curve is shown in the top panel of Fig.\,\ref{fig:CoRoT}. Its Rayleigh frequency resolution amounts to $\delta f_{\rm ray}{=}1/150.49{=}0.00665$\,\d).

\subsection{Iterative prewhitening}

\begin{table}[t]
  \caption{Significant frequencies in the CoRoT light curve of HD\,43317 ordered and labelled by increasing frequency.}
\centering
\tabcolsep=6pt
\resizebox{\columnwidth}{!}{
\begin{tabular}{llllll}
\hline
\hline
		& Frequency 		& Formal $\delta f$ 	& Amplitude & S/N	& Identification\\
		& Value		  	& from NLLS			& [ppm]			\\
		& [\d]			& [$10^{-4}$\d]		& $\pm 1$		\\
\hline
$f_{	1	}$&	0.22323	& 0.14 &	228	&	8.8	&		\\
$f_{	2	}$&	0.39059	& 0.39 &	83	&	4.5	&		\\
$f_{	3	}$&	0.52900	& 0.& 98	&	5.0	&		\\
$f_{	4	}$&	0.69162	& 0.50 &	65	&	4.2	&	g mode	\\
$f_{	5	}$&	0.75289	& 0.23 &	140	&	6.2	&	g mode	\\
$f_{	6	}$&	0.82781	& 0.41 &	80	&	4.7	&	g mode	\\
$f_{	7	}$&	0.87523	& 0.28 &	115	&	5.2	&	g mode	\\
$f_{	8	}$&	0.92786	& 0.24 &	135	&	5.8	&	g mode	\\
$f_{	9	}$&	0.99541	& 0.32 &	1013	&	20.7	&	g mode	\\
$f_{	10	}$&	1.07137	& 0.24 &	136	&	6.0	&		\\
$f_{	11	}$&	1.10039	& 0.02 &	1456	&	21.3	&	g mode	\\
$f_{	12	}$&	1.11457	& 0.18 &	196	&	6.3	&	$f_{\rm rot}$	\\
$f_{	13	}$&	1.17536	& 0.05 &	672	&	16.9	&	g mode	\\
$f_{	14	}$&	1.22798	& 0.39 &	82	&	4.9	&	g mode	\\
$f_{	15	}$&	1.34244	& 0.17 &	209	&	7.2	&	g mode	\\
$f_{	16	}$&	1.35291	& 0.40 &	86	&	4.9	&	g mode	\\
$f_{	17	}$&	1.50345	& 0.20&	165	&	8.3	&		\\
$f_{	18	}$&	1.64825	& 0.43 &	75	&	5.2	&	$2f_{6}$	\\
$f_{	19	}$&	1.68418	& 0.08 &	427	&	13.5	&		\\
$f_{	20	}$&	1.70454	& 0.14 &	232	&	8.6	&	g mode	\\
$f_{	21	}$&	1.81560	& 0.07 &	457	&	11.8	&	g mode	\\
$f_{	22	}$&	2.09539	& 0.19 &	168	&	7.9	&	$f_{9} + f_{11}$	\\
$f_{	23	}$&	2.15501	& 0.25 &	131	&	9.2	&		\\
$f_{	24	}$&	2.17658	& 0.21 &	155	&	8.7	&	$f_{9} + f_{13}$	\\
$f_{	25	}$&	2.22787	& 0.24 &139	&	9.5	&	$2f_{\rm rot}$	\\
$f_{	26	}$&	2.26314	& 0.21 &	156	&	8.8	&		\\
$f_{	27	}$&	2.74964	& 0.29 &	112	&	10.2	&		\\
$f_{	28	}$&	3.31881	& 0.27 &	121	&	11.0	&	$3f_{\rm rot}$	\\
$f_{	29	}$&	3.49578	& 0.26 &	125	&	10.4	&	g mode	\\
$f_{	30	}$&	3.57826	& 0.40 &	82	&	9.6	&		\\
$f_{	31	}$&	4.33105	& 0.06 &	557	&	34.0	&	g mode	\\
$f_{	32	}$&	5.00466	& 0.45 &	72	&	8.2	&	g mode	\\
$f_{	33	}$&	5.02538	& 0.50 &	66	&	8.1	&		\\
$f_{	34	}$&	5.49930	& 0.34 &	95	&	11.1	&	$2f_{27}$	\\
$f_{	35	}$&	6.35786	& 0.45 &	71	&	12.5	&		\\
\hline
\end{tabular}}
\label{tab:frequency}
\tablefoot{For each frequency, we provided the amplitude and its signal-to-noise level (S/N) after the iterative prewhitening.  We also provided the formal errors computed by a non-linear least-squares (NLLS).  The global amplitude error is approximate and was not used in the forward modelling.  When applicable, we indicate the identification of the frequency following Sect.\,\ref{sec:freq_cleaning} and \ref{sec:modelling}.}
\end{table}

We utilized the corrected CoRoT light curve to determine the frequencies and amplitudes of the periodic photometric variability by means of iterative prewhitening.  This approach is appropriate for intermediate- and high-mass pulsators \citep[e.g.,][]{2009A+A...506..111D}.  To deduce the significance of a frequency in the ten times oversampled Lomb-Scargle periodogram \citep{1976Ap+SS..39..447L, 1982ApJ...263..835S}, we relied on its signal-to-noise (S/N) level. A frequency was accepted if it had a S/N$ \geq 4$ in amplitude \citep{1993A+A...271..482B}, where the noise level was computed in a frequency window of 1\,\d centered at the frequency peak after its corresponding variability was prewhitened.  It has been demonstrated \citep[see e.g.,][]{2009A+A...506..111D, 2012A+A...542A..55P, 2012A+A...542A..88D} that such a S/N criterion is more appropriate over the false alarm probability (i.e., p-criterion) due to the correlated nature and the noise properties of CoRoT data.  The narrow frequency window of 1\,\d was chosen over wider frequency windows to have a conservative frequency list for the subsequent modelling procedure.

We performed the iterative prewhitening in the frequency domain spanning from 0 -- 10\,\d.  No significant pulsation frequencies occurred at higher values, and aliasing structure due to the satellite-orbit frequency at $f_{\rm sat} = 13.97$\,\d is not of interest for the interpretation of the modes.  In total, we deduced 35 significant frequencies with values ranging from 0.2232\,\d up to 6.3579\,\d.  We provide the optimised frequency and amplitude values after non-linear least-squares fitting in the time domain in Table\,\ref{tab:frequency}, as well as their corresponding S/N values.  We also mark these frequencies in the periodogram of the CoRoT light curve, together with the periodogram of the residual light curve, in Fig.\,\ref{fig:CoRoT}.  The formal frequency and amplitude errors following the method of \citet{1999DSSN...13...28M} were calculated, and these formal errors are also listed in Table\,\ref{tab:frequency}.  We recall that amplitudes were not used in the subsequent in the modelling.  It is well known that such formal error estimates grossly underestimate the true errors, because of the simplistic assumption of uncorrelated data with white Gaussian noise. This underestimation is particularly prominent in the case of CoRoT data for g-mode pulsators, as these are highly correlated in nature and are subject to heteroscedastic errors \citep[see][]{2009A+A...506..111D}.  Correcting for these two properties is far from trivial but leads to at least one and possibly two orders of magnitude increase for the frequency errors given in Table\,\ref{tab:frequency}.

\subsection{Selecting the individual mode frequencies}
\label{sec:freq_cleaning}
Forward seismic modelling is commonly done by fitting the frequencies of independent pulsation modes. Hence, the frequencies extracted from the CoRoT photometry of HD\,43317 by the iterative prewhitening needed to be filtered for combination frequencies.  For this reason, we limited the frequency list to values that could not be explained by frequency harmonics or low-order (2 or 3) linear combinations of the 5 highest amplitude frequency peaks \citep[following the method of e.g.,][]{2015MNRAS.450.3015K, 2017ampm.book.....B}.  Low-order combination or harmonic frequencies typically have a smaller amplitude than the individual parent mode frequencies.  Thus, we identified a frequency as a combination frequency if it had a smaller amplitude than the parent frequencies.  We also excluded frequencies that corresponded to the known rotation frequency of the star, $f_{\rm rot} = 1.113991(5)$\,\d \citep{2017A+A...605A.104B}, and its harmonics.

By using these various criteria, we removed a total of seven frequencies from the list (see Table\,\ref{tab:frequency}) and identified the remaining 28 as candidate pulsation mode frequencies for the forward seismic modelling.

\section{Forward seismic modelling}
\label{sec:modelling}
\subsection{Setup}

To determine the stellar structure of HD\,43317, we computed a grid of non-rotating, non-magnetic 1D stellar structure and evolution models employing MESA \citep[v8118,][]{2011ApJS..192....3P, 2013ApJS..208....4P, 2015ApJS..220...15P}.  For each MESA model in the grid, we computed linear adiabatic pulsation mode frequencies of dipole and quadrupole geometries using the pulsation code GYRE \citep[v4.1,][]{2013MNRAS.435.3406T}.  These theoretical predictions for the pulsation mode frequencies were then quantitatively compared to the frequencies extracted from the CoRoT light curve to deduce the best fitting MESA models.  Such a grid-based modelling approach has been successfully used earlier to interpret the pulsation mode frequencies of g-mode pulsations in rotating stars based on {\it Kepler\/} space photometry (e.g., \citealt{2016A+A...593A.120V} for $\gamma$\,Dor pulsators and \citealt{2016ApJ...823..130M} for SPB stars).  A good grid set up requires appropriate evaluation of the effect of the choices for the input physics of stellar models on the predicted pulsation frequencies. For g modes in stars with a convective core, a global assessment for non-magnetic pulsators was recently offered by Aerts et al.\,(submitted). These authors showed that typical uncertainties for the theoretical predictions of low-degree g-mode pulsation frequencies range from 0.001 to 0.01\d, depending on the specific aspect of the input physics that is being varied.  The grid of MESA models in this paper was constructed accordingly.

Compared to the forward modelling of rotating g-mode pulsators based on {\it Kepler\/} space photometry, the case of HD\,43317 is hampered by two major limitations: i) the ten times poorer frequency resolution of the data that led to fewer significant frequencies and prevented mode identification from period spacing patterns; ii) the unknown effect of a magnetic field in the theoretical prediction of the pulsation frequencies. For this reason, the MESA grid was limited to the minimum number of free stellar parameters to be estimated for meaningful seismic modelling: the stellar mass $M_{\star}$, the central hydrogen mass fraction $X_c$ (as a good proxy for the age), and the convective core overshooting. For the latter, we used an exponential overshooting description with parameter $f_{\rm ov}$. Each of these three parameters were allowed to vary in a sensible range appropriate for B-type g-mode pulsators and using discrete steps.  The stellar mass ranged from 4.0\,$\mathrm{M_{\odot}}$ to 8.0\,$\mathrm{M_{\odot}}$, with a step of 0.5\,$\mathrm{M_{\odot}}$.  For each of these masses, MESA models were evolved from the zero-age main-sequence (ZAMS), corresponding to $X_c{\sim}0.70$, to the terminal-age main-sequence (TAMS), defined as $X_c{=}0.001$; we saved the stellar structure models at discrete $X_c$ steps of $0.01$.  The parameter for the exponential convective core overshooting, $f_{\rm ov}$, is a dimensionless quantity expressed in units of the local pressure scale height; it was varied from 0.002 up to 0.040, with a step of 0.002.  We refer to \citet{1996A+A...313..497F}, \citet{2000A+A...360..952H} and \citet{2018arXiv180202051P} for explicit descriptions of this formulation of core overshooting.  We followed the method of \citet{2018arXiv180202051P} to set the exponential overshooting in MESA, where we employed a $f_0 = 0.005$.  All other aspects of the input physics were kept fixed according to the MESA inlist provided in Appendix\,\ref{sec:appendix_MESA}, following the guidelines in Aerts et al.\ (submitted). In particular, a solar initial metallicity was adopted according to the spectroscopic results based on a NLTE abundance analysis by \citet{2012A+A...542A..55P}.  The chemical mixture was taken to be the solar one from \citet{2009ARA+A..47..481A}. We used the opacity tables from \citet{2016MNRAS.455L..67M}.  Further, we relied on the Ledoux criterion to determine the convection boundaries and adopted the mixing length theory by \citet{1968pss..book.....C}, with mixing length parameter $\alpha_{\rm mlt}{=}2.0$ based on the solar calibration by \citet{1996Sci...272.1286C}.  We fixed the semi-convection parameter $\alpha_{\rm sc}{=}0.01$ and included additional constant diffusive mixing in the radiative region $D_{\rm ext}{=}10$\,cm$^2$\,s$^{-1}$ following \citet{2016ApJ...823..130M}.  The MESA computations were started from pre-computed pre-main-sequence models that come with the installation suite of the code.

For each of the MESA models in the grid, we computed the adiabatic pulsation frequencies for dipole ($\ell{=}1$) and quadrupole ($\ell{=}2$) g modes with GYRE, given that recent {\it Kepler\/} data of numerous rotating F-and B-type g-mode pulsators all revealed such low-degree modes \citep[e.g.,][]{2016A+A...593A.120V, 2017A+A...598A..74P, 2017MNRAS.467.3864S}.  The effects of rotation on the pulsation mode frequencies were included in the GYRE calculations by enabling the traditional approximation \citep[see e.g.,][]{1960..book.....E, 2003MNRAS.343..125T}.  While doing so, we assumed that the star is a rigid rotator with $f_{\rm rot}{=}1.113991$\,\d, as determined from the magnetometric analysis (at the stellar surface) by \citet{2017A+A...605A.104B} and following the findings by \citet{2017ApJ...847L...7A} and Van Reeth et al.\,(submitted).  We computed all dipole and quadrupole g modes with radial orders from $n_g{=}-1$ to $n_g{=}-75$.  The settings for the GYRE computations are contained in Appendix\,\ref{sec:appendix_GYRE}.

We adopted a quantitative frequency matching approach and compared the GYRE pulsation mode frequencies with the observed CoRoT frequencies, as is common practice in forward seismic modelling.  As such, we determined the closest model frequency to an observed frequency, and used these to compute the reduced $\chi^2$ value of the fit.  While doing so, any predicted model frequency was allowed to match only one observed frequency.  The reduced $\chi^2$ values were defined as: 
\begin{equation}
\chi^2 = \frac{1}{N - k - 1} \sum\limits_{i=1}^{N}\left( \frac{f_{\rm obs}^i - 
f_{\rm mod}^i}{\delta f_{\rm ray}} \right)^2 \,\mathrm{,}
\label{eq:chi_square}
\end{equation}
\noindent with $N>4$ the total number of frequencies included in the fit, $k=3$ the number of free parameters to estimate (i.e., the three input parameters of the MESA grid), $f_{\rm obs}^i$ the considered detected frequency in the CoRoT data, $f_{\rm mod}^i$ the theoretically predicted GYRE pulsation mode frequency, and $\delta f_{\rm ray}$ the Rayleigh frequency resolution of the CoRoT light curve.  The square of the latter was taken to be a good overall estimate of the variance that encapsulated the frequency resolution of the CoRoT data and the errors on the theoretical frequency predictions by GYRE, given that the formal errors in Table\,\ref{tab:frequency} were unrealistically small and the theoretical predictions were of similar order of magnitude than the Rayleigh limit for the data set of HD\,43317. Moreover, due to lack of an {\it a priori\/} mode identification and evolutionary status of the star, it was recommendable to give equal weight to each of the detected frequencies in the fitting procedure because the mode density varies considerably during the evolution along the main sequence.  We come discuss this issue further in the following sections.

During the forward modelling of HD\,43317, the $\chi^2$ value of the fit to the GYRE frequencies of a given MESA model was not the only diagnostic.  We also considered the location of the best MESA models (i.e., those with the lowest $\chi^2$ value) in the Kiel diagram to investigate how unique the best solution was.  In case the parameters of the MESA models were well constrained by the fitting process, we anticipated the best models to lie within the same location in the Kiel diagram representing a unique $\chi^2$ valley in the 3D parameter space of the MESA grid, which then allowed us to create simplistic confidence intervals for the free parameters, as explained below.  This method was a necessary complication compared to the modelling of {\it Kepler\/} stars because no obvious independent period spacing pattern spanning several consecutive radial orders could be identified for HD\,43317. Hence the straightforward method for mode identification using the technique employed by \citet{2015A+A...574A..17V, 2016A+A...593A.120V} was not possible for our CoRoT target star.

\subsection{Blind forward modelling}
\label{sec:modelling_blind}

\begin{figure}[t]
		\centering
			\includegraphics[width=0.45\textwidth, height = 0.33\textheight]{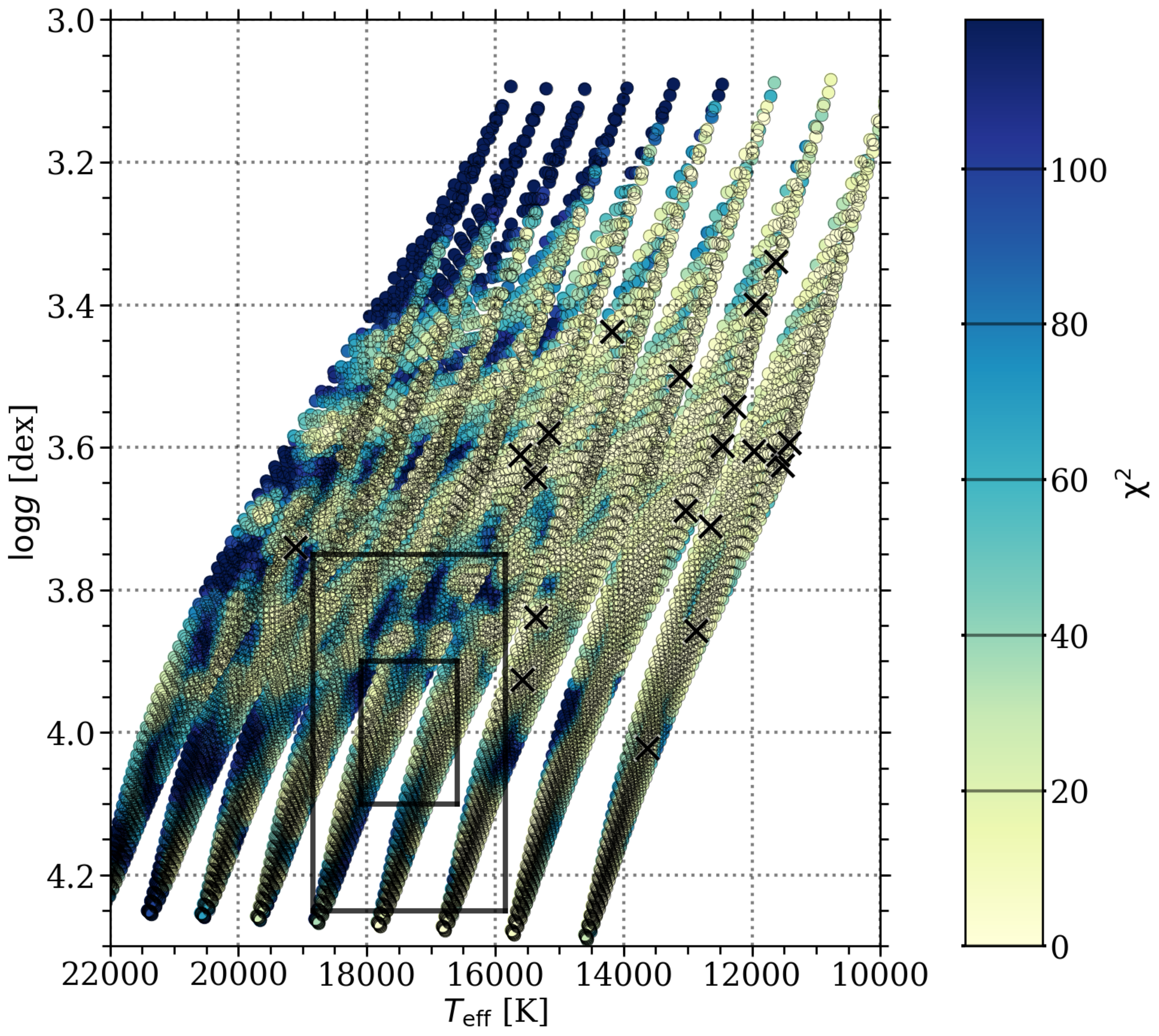}%
\caption{Kiel diagram showing the results of blind forward modelling based on all 28 pulsation mode frequencies.  The colour scale indicates  the $\chi^2$ level and the black crosses represent the location of the 20 best fitting MESA models.  The $1\sigma$ and $2\sigma$ confidence intervals for the derived spectroscopic parameters by \citet{2012A+A...542A..55P} are shown by the black boxes.}
\label{fig:initial_GYRE}
\end{figure}

Although the CoRoT data were sufficient to detect g modes in B-type stars, the relatively large Rayleigh frequency resolution did not allow for unambiguous mode identification in the case of HD\,43317 due to its fast rotation. This is different from the case of the CoRoT ultra-slow rotator HD\,50230, whose g-mode periods constituted the first discovered period spacing pattern that led to mode identification for a main-sequence star thanks to the absence of a slope in its pattern \citep{2010Natur.464..259D}.  However, although not available for HD\,43317,  the ten times longer light curves observed by the {\it Kepler\/} mission for tens of g-mode pulsators meanwhile provided critical information on the types of modes expected in such rapidly rotating pulsators \citep[e.g.,][for sample papers with B- and F-type stars]{2015A+A...574A..17V, 2017A+A...598A..74P, 2017MNRAS.467.3864S}.  This led these authors to the conclusion that almost all detected g modes were dipole or quadrupole modes, irrespective of the rotation rate of the g-mode pulsator.  We relied on this knowledge from {\it Kepler\/} to guide and perform our modelling of HD\,43317.

\begin{figure*}[t]
\centering\includegraphics[width=\textwidth, height = 0.50\textheight]{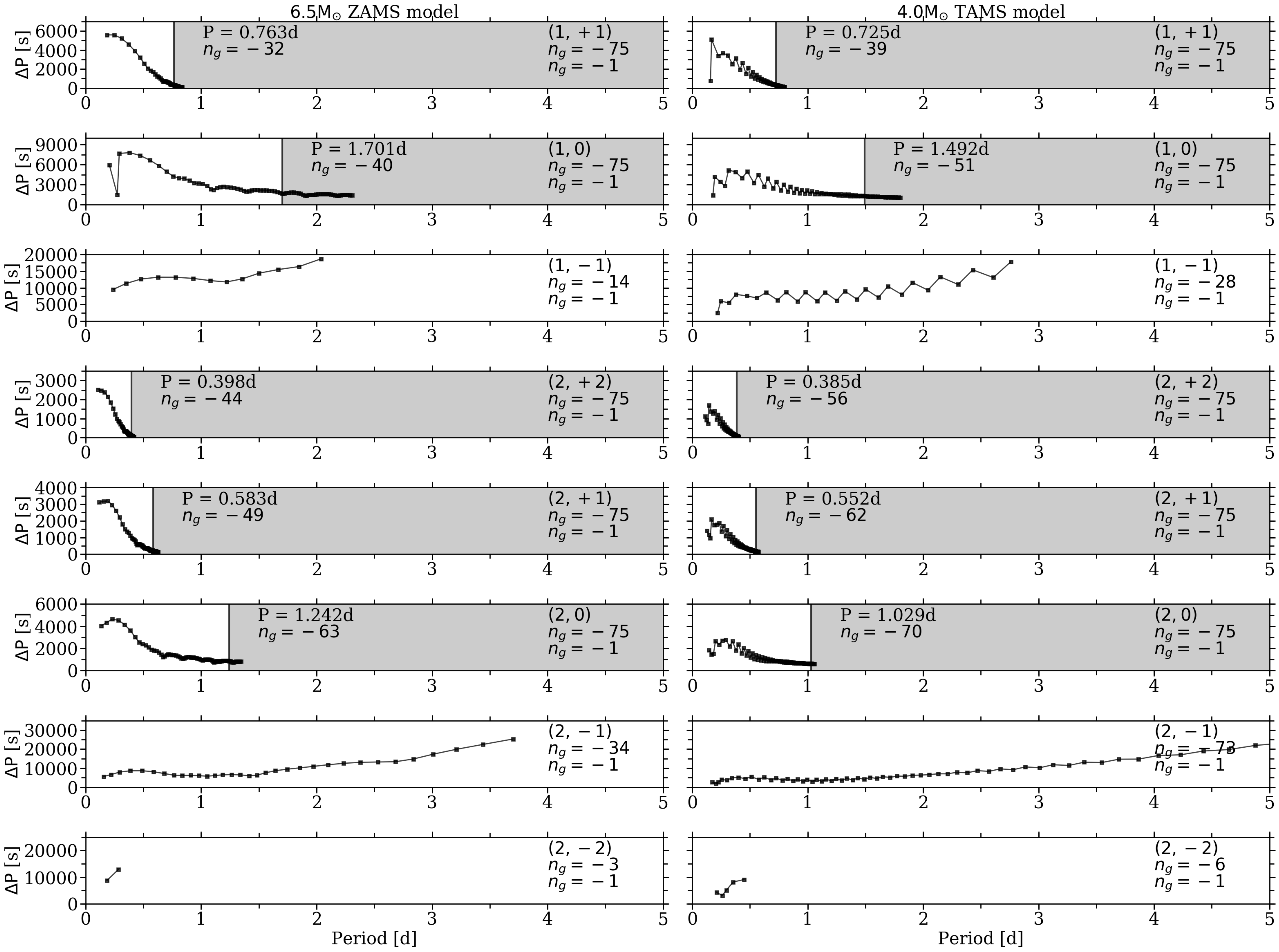}%
\caption{Period spacing patterns, as computed by GYRE and accounting for rigid rotation at the rate of HD\,43317, using the traditional approximation, for two demonstrative MESA models.  \textit{Left:} a 6.5\,$\mathrm{M_{\odot}}$ MESA model at the ZAMS; \textit{right:} a 4.0\,$\mathrm{M_{\odot}}$ model close to the TAMS.  Each row represents the theoretical period spacing pattern for a given mode geometry, with the range of radial orders as listed.  Grey boxes correspond to regions where the frequency difference between theoretical frequencies of modes with consecutive radial order is smaller than the Rayleigh frequency resolution of the CoRoT data; these ranges thus cannot be used. The corresponding pulsation period and radial order for the limiting theoretical mode to be used in the modelling for that given MESA model and mode geometry is listed in each upper left corner.}
\label{fig:GYRE_allmodes}
\end{figure*}

Without a period spacing pattern to identify (some of) the 28 candidate pulsation mode frequencies, we first permitted the observed frequencies to match with any of the theoretical frequencies of the predicted dipole or quadrupole mode geometries for a given MESA model, without taking the constraints from spectroscopy into account.  The locations of the 20 models with the lowest $\chi^2$ value resulting from such ``blind'' modelling in the Kiel diagram are shown in Fig.\,\ref{fig:initial_GYRE}.  These models corresponded to various stellar masses, spanning the entire MESA grid.  Moreover, the lowest $\chi^2<1$ were reached for models with a low $X_c$ value, corresponding with the lower $\log g$ values in the grid.  This result is a common feature of such unconstrained forward modelling without any restriction on the identification of the modes or on the evolutionary stage. Indeed, the mode density in the frequency region of interest is much higher for models near the TAMS than for less evolved stages (see Fig.\,\ref{fig:GYRE_allmodes}).  Hence, the $\chi^2$ values will always be artificially smaller for evolved models near the TAMS than for less evolved models in absence of mode identification.  Moreover, the majority of these best models relied on zonal mode frequencies, while many of these are unresolved in the CoRoT data (see Fig.\,\ref{fig:GYRE_allmodes}).

This exercise illustrated that blind forward modelling without spectroscopic constraints is not a good strategy in the absence of mode identification. Indeed, the spectroscopic parameters of these best models did not agree with the observationally derived $2\sigma$ confidence intervals by \citet{2012A+A...542A..55P}, i.e., $T_{\mathrm{eff}} = 17350\pm1500$\,K and $\log g{=}4.0 \pm 0.25$\,dex.  Therefore, a dedicated strategy for the forward seismic modelling of HD\,43317 was required.

\subsection{Conditional forward modelling}
\label{sec:modelling_hypothesis}
\begin{figure*}[t]
\centering
\includegraphics[width=\textwidth, height = 0.66\textheight]{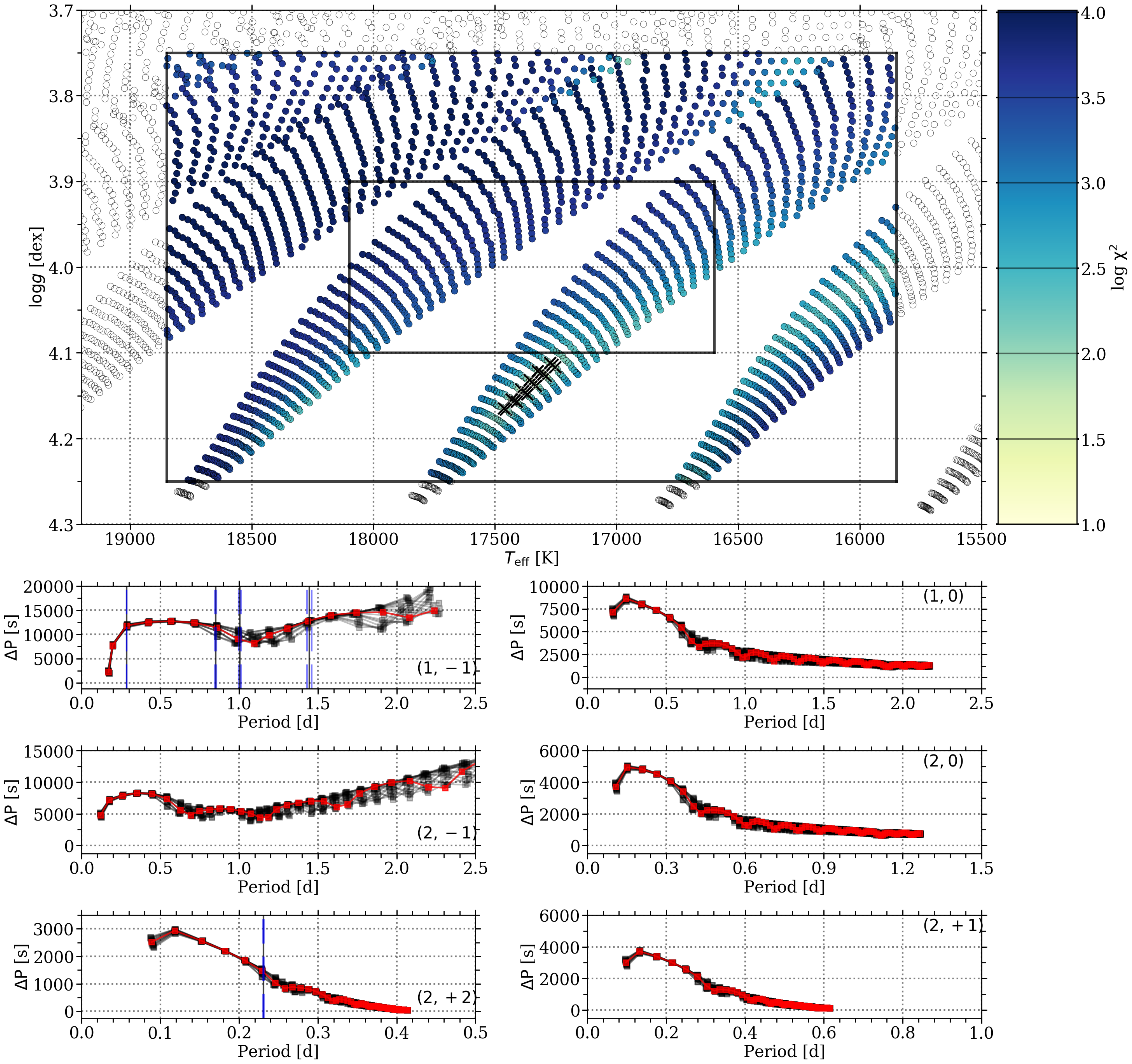}%
\caption{Summary plot for the results of the forward modelling of HD\,43317 based on the dipole mode hypothesis (see text).  \textit{Top}: Kiel diagram showing the location of the MESA models whose GYRE frequencies come closest to the detected frequencies.  The color scale represents the $\chi^2$ value (see Eq.\,(\ref{eq:chi_square})), the open symbols indicate MESA models outside the spectroscopic limits of the star deduced by \citet[][with the $1\sigma$ and $2\sigma$ limits indicated by the black boxes]{2012A+A...542A..55P}, and the black crosses represent the location of the 20 best fitting MESA models.  \textit{Bottom panels}: period spacing patterns of the 20 best fitting MESA models for various mode geometries (black lines and squares).  Those of the best description are shown in red.  The observed pulsation mode periods used in the $\chi^2$ are indicated by the vertical grey lines, while the vertical blue regions indicate the observed period range taking into account the Rayleigh limit of the data.  Only $(1,-1)$ modes (i.e., $f_{4}$, $f_{9}$, $f_{13}$, and $f_{29}$, as well as the $f_{31}$ as a $(2,+2)$ mode) were relied upon to achieve the frequency matching.}
\label{fig:dipole_GYRE}
\end{figure*}

We adopted various hypotheses in the best model selection for the estimation of the three stellar parameters $(\mathrm{M_{\star}}, X_c, f_{\rm ov})$.  The first hypothesis was that any selected MESA model should comply with the spectroscopic properties of the star.  Thus, the effective temperature, $T_{\mathrm{eff}}$, and the surface gravity, $\log g$, were demanded to be consistent with those derived from high-precision spectroscopy.  As such, we excluded any MESA models that did not agree with the $2\sigma$ confidence intervals of the combined spectroscopic analysis done by \citet{2012A+A...542A..55P}.  This assumption resulted in the exclusion of MESA models near the TAMS, leaving only models with comparable mode density.

The next assumption was that any appropriate MESA model should be able to explain several of the highest-amplitude frequencies as pulsation mode frequencies.  Indeed, it did not make sense to accept models that explained low-amplitude modes while they did not lead to a good description of the dominant modes.  We (arbitrarily) placed the cut-off limit at an amplitude of 500\,ppm, thus requiring that $f_{9}$, $f_{11}$, $f_{13}$, and $f_{31}$ were matched.  Using spectroscopic mode identification techniques, \citet{2012A+A...542A..55P} identified $f_{31}$ to be a prograde $(\ell, m){=}(2,+2)$ mode\footnote{We defined modes with $m > 0$ as prograde modes and $m < 0$ as retrograde modes.}.  We thus fixed the mode identification of this particular frequency during the modelling process.  The other three dominant modes in the CoRoT photometry were not detected in the spectroscopy.  This is not so surprising, given that photometric and spectroscopic diagnostics are differently affected by mode geometries \citep[see Chapter\,6 in][]{2010aste.book.....A} and that we have no information on the intrinsic amplitudes of excited modes.  Hence, we were not able to use the identification of $f_{31}$ to place constraints on the degree of the other highest-amplitude modes.  However, given the dominance of a sectoral mode in the spectroscopy, we excluded a pole-on view of the star, because it would correspond to an angle of complete cancellation for such a mode \citep[at $i=0\,^{\circ}$; see e.g.,][]{2010aste.book.....A}.  As such, \citet{2012A+A...542A..55P} constrained the inclination angle to be $i\in [20^\circ, 50^\circ]$.

We considered different hypotheses on the identification of the modes, as described in the following sections.  These hypotheses followed from comparison of the theoretical period spacing patterns with those from the detected frequencies (comparing Table\,\ref{tab:frequency} to Fig.\,\ref{fig:GYRE_allmodes}).  For most of the $m > 0$ modes, the frequency difference between modes of consecutive radial order was smaller or comparable to the Rayleigh limit of the CoRoT data.  The same is true for most zonal modes, especially for the $(\ell, m){=}(2,0)$ modes.  As such, creating a period spacing pattern spanning consecutive radial orders for such modes from the observed frequencies was not meaningful with the frequency resolution of the CoRoT data.  Therefore, we assumed that the observed frequencies were $m=-1$ modes, unless explicit evidence was available (e.g., for $f_{31}$ from spectroscopy). This was a valid assumption considered the results of \textit{Kepler} for g-mode pulsators \citep{2016A+A...593A.120V} and the measured rotation rate of HD\,43317.  We then included additional observed frequencies under the made hypotheses.

With this piece-wise conditional modelling, we intended to constrain both the parameters of HD\,43317 and the mode geometry of the detected frequencies.  The theoretical period spacing patterns of the best models at any given step denoted which observed frequency could be additionally included in the frequency matching.  Detected frequencies were considered for inclusion  if they were sufficiently close to a theoretical frequency predicted by the best models.  Ideally, this modelling scheme should lead to a well clustered group of MESA models in the Kiel diagram for a large number of the 28 detected frequencies, leading to an estimate on their mode geometry.  To ensure robustness and reproducibility of a hypothesis during the forward seismic modelling, we defined the following criteria for any good solution:
\begin{itemize}
\item At least five detected frequencies must be accounted for in the use of Eq.\,(\ref{eq:chi_square}).
\item The location of the 20 best MESA models in the Kiel diagram must remain consistent with the $2\sigma$ spectroscopic error box.
\end{itemize}

\subsection{Dipole retrograde mode hypothesis}
\label{sec:modelling_dipole}

\begin{figure*}[t]
\centering
\includegraphics[width=\textwidth, height = 0.66\textheight]{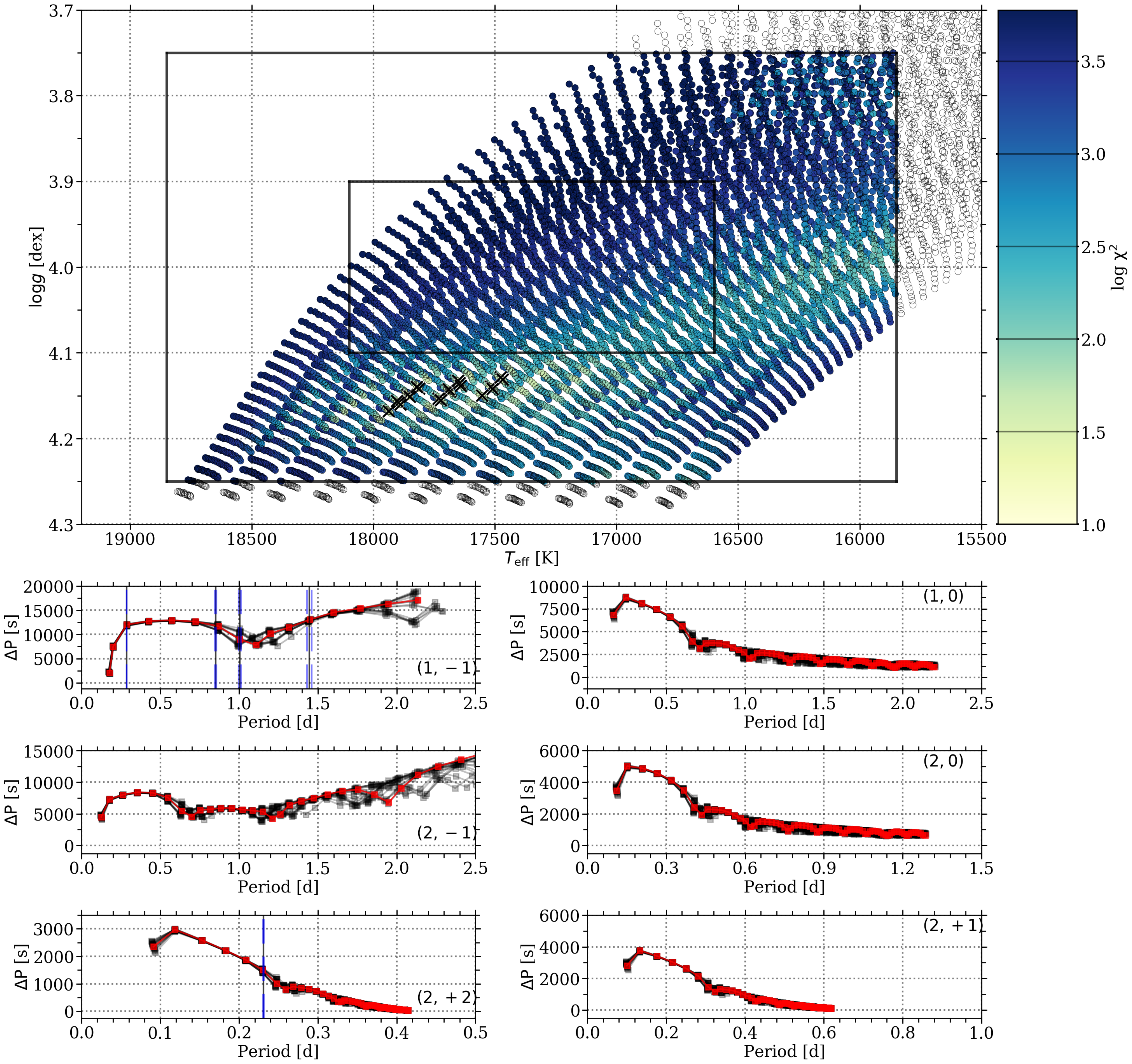}%
\caption{Summary plot for the overall forward modelling of HD\,43317 using the dipole mode hypothesis on the refined grid of MESA models.  The same colour scheme as Fig.\,\ref{fig:dipole_GYRE} is used.}
\label{fig:dipole_GYRE_fine}
\end{figure*}

\begin{figure*}[t]
\centering
\includegraphics[width=\textwidth, height = 0.66\textheight]{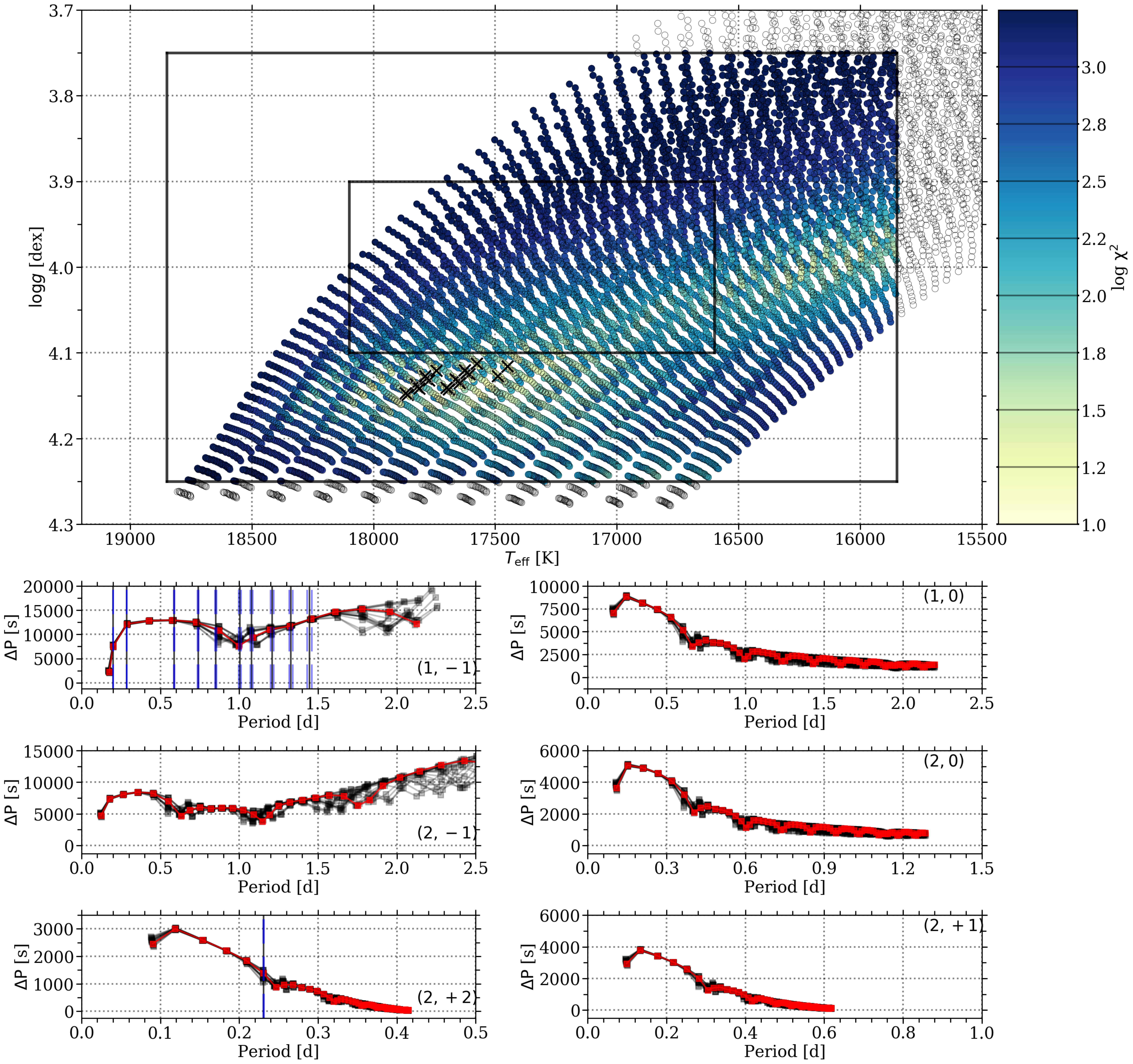}%
\caption{Summary plot for the overall forward modelling of HD\,43317 using the extended dipole mode hypothesis on the refined grid of MESA models.  The same colour scheme as Fig.\,\ref{fig:dipole_GYRE} is used.}
\label{fig:dipole_GYRE_extended}
\end{figure*}

As a first hypothesis, we considered the case that the three dominant pulsation mode frequencies $f_{9}$, $f_{11}$, $f_{13}$ corresponded to dipole retrograde modes.  Dipole modes were by far the most frequently detected ones in space photometry of g-mode pulsators \citep[e.g.,][]{2014A+A...570A...8P, 2016ApJ...823..130M, 2016A+A...593A.120V}.  At the measured $v\sin\,i$ and $f_{\rm rot}$ of HD\,43317, most of the detected g-mode frequencies in pulsators with {\it Kepler\/} photometry would belong to a series of $(1,-1)$ modes \citep{2016A+A...593A.120V}.  As such, it was reasonable to assume that $f_{9}$, $f_{11}$, $f_{13}$ were $(1,-1)$ modes.

Using these three frequencies and $f_{31}$ constrained by spectroscopy,the forward modelling led to models with a mass ranging from 5.0\,$\mathrm{M_{\odot}}$ up to 6.5\,$\mathrm{M_{\odot}}$.  A comparison of the observed frequencies with the model frequencies suggested that $f_{11}$ was likely not a $(1,-1)$ mode.  We thus dropped the requirement on this frequency and repeated the forward modelling with only $f_{9}$ and $f_{13}$ as $(\ell, m){=}(1,-1)$ modes and $f_{31}$ as a $(2,+2)$ mode.  This returned two families of solutions in the Kiel diagram, namely a group with $M_{\star}{=}5.0$\,$\mathrm{M_{\odot}}$ with $X_c{\approx}0.50$, and a stripe of models with a mass of 5.5\,$\mathrm{M_{\odot}}$, while $X_c$ decreases from 0.63 to 0.52 and $f_{\rm ov}$ from 0.038 to 0.022.  All these models resulted in similar theoretical $(1,-1)$ frequencies and identified $f_{29}$ to have this mode geometry as well.  We thus included this observed frequency in the hypothesis and repeated the forward modelling.

The modelling converged to only one family of solutions in the Kiel diagram that corresponded with $M_{\star}{=}5.5$\,$\mathrm{M_{\odot}}$, $X_c = 0.53 - 0.58$ and $f_{\rm ov}$ ranging from 0.016 to 0.022.  These models indicated that the observed frequencies $f_{3}$, $f_{4}$, $f_{5}$, $f_{6}$, $f_{16}$ and $f_{20}$ should be part of the $(1,-1)$ period spacing pattern.  We included only $f_{4}$, because it was the closest match and repeated the modelling with the updated hypothesis.

We recovered the same set of 20 best MESA models with $f_{4}$ included in the hypothesis and noted that five additional observed frequencies agreed with the theoretical predictions.  Hence, we satisfied our defined criteria.  The conclusion of the modelling with this hypothesis is represented in Fig.\,\ref{fig:dipole_GYRE}.  The top panel shows the Kiel diagram with the location of the 20 best MESA models, as well as the $\chi^2$ values of the fitting process to the theoretical GYRE frequencies.  The model period spacing patterns of the 20 best fitted models (and the corresponding observed frequencies in the assumption) are given in the bottom panels of Fig. \ref{fig:dipole_GYRE}.

\subsection{Increasing the mass resolution of the MESA grid}
\label{sec:modelling_quadrupole}
To investigate the robustness of our obtained solution shown in Fig.\,\ref{fig:dipole_GYRE}, we created a new grid of MESA models that had a finer mass resolution.  This new grid had the same settings for the micro- and macro-physics as before, but the stellar masses ranged from 5.0\,$\mathrm{M_{\odot}}$ up to 6.0\,$\mathrm{M_{\odot}}$ with a step size of 0.1\,$\mathrm{M_{\odot}}$.  We repeated the frequency matching of the five observed frequencies to the GYRE frequencies of the finer grid of models and summarized the result in Fig.\,\ref{fig:dipole_GYRE_fine}.  The 20 best MESA models corresponded to slightly different values for the stellar mass and exponential overshooting factor. These values were unavailable in the coarser grid.

The theoretical frequencies for the already identified $(1,-1)$ modes of the now 20 best models did not alter appreciably compared to those of the solution in the coarse grid, nor did the radial orders change.  We included additional identifications as dipole modes, namely $f_{5}$, $f_{6}$, $f_{8}$, $f_{16}$, $f_{20}$ and $f_{32}$.  The result of the modelling with these ten $(1,-1)$ and one $(2, +2)$ mode is illustrated in Fig.\,\ref{fig:dipole_GYRE_extended}.

We noted that the location of the 20 best MESA models now moved to slightly lower $X_c$ values when using this more extended frequency list.  However, the confidence intervals of these solutions for the different hypotheses still overlapped at a $2\sigma$ level, i.e., the location of the best solutions did not move outside the range of their variance. Here, the $\alpha$\,\% confidence interval is defined by the upper limit on the $\chi^2$ value as
\begin{equation}
\chi^2_{\alpha} = \frac{\chi^2_{\alpha, k} \cdot \chi^2_{\rm min}}{k}  \, \mathrm{,}
\label{eq:upperlimit}
\end{equation}
\noindent where $\chi^2_{\rm min}$ is the $\chi^2$ value for the best fit and $\chi^2_{\alpha, k}$ the tabulated value for an $\alpha$\,\% inclusion of the cumulative distribution function of a $\chi^2$ distribution with $k=3$ degrees of freedom.

A local minimum in the $\chi^2$ landscape occurred for models with a stellar mass of 5.3\,$\mathrm{M_{\odot}}$ and $X_c \approx 0.40$ (i.e., $T_{\rm eff} = 16000$\,K and $\log g = 4.0$\,dex), but their $\chi^2$ values were larger than the upper limit of the $2\sigma$ confidence interval of the best solutions in the minimum dictated by the spectroscopic limits.  This family of second-best models were only compatible at a $3\sigma$ level (cf.\ Fig.\,\ref{fig:combined_chisquare}.)

\subsection{Adding modes with other mode geometry}

\begin{figure*}[t]
\centering
\includegraphics[width=\textwidth, height = 0.66\textheight]{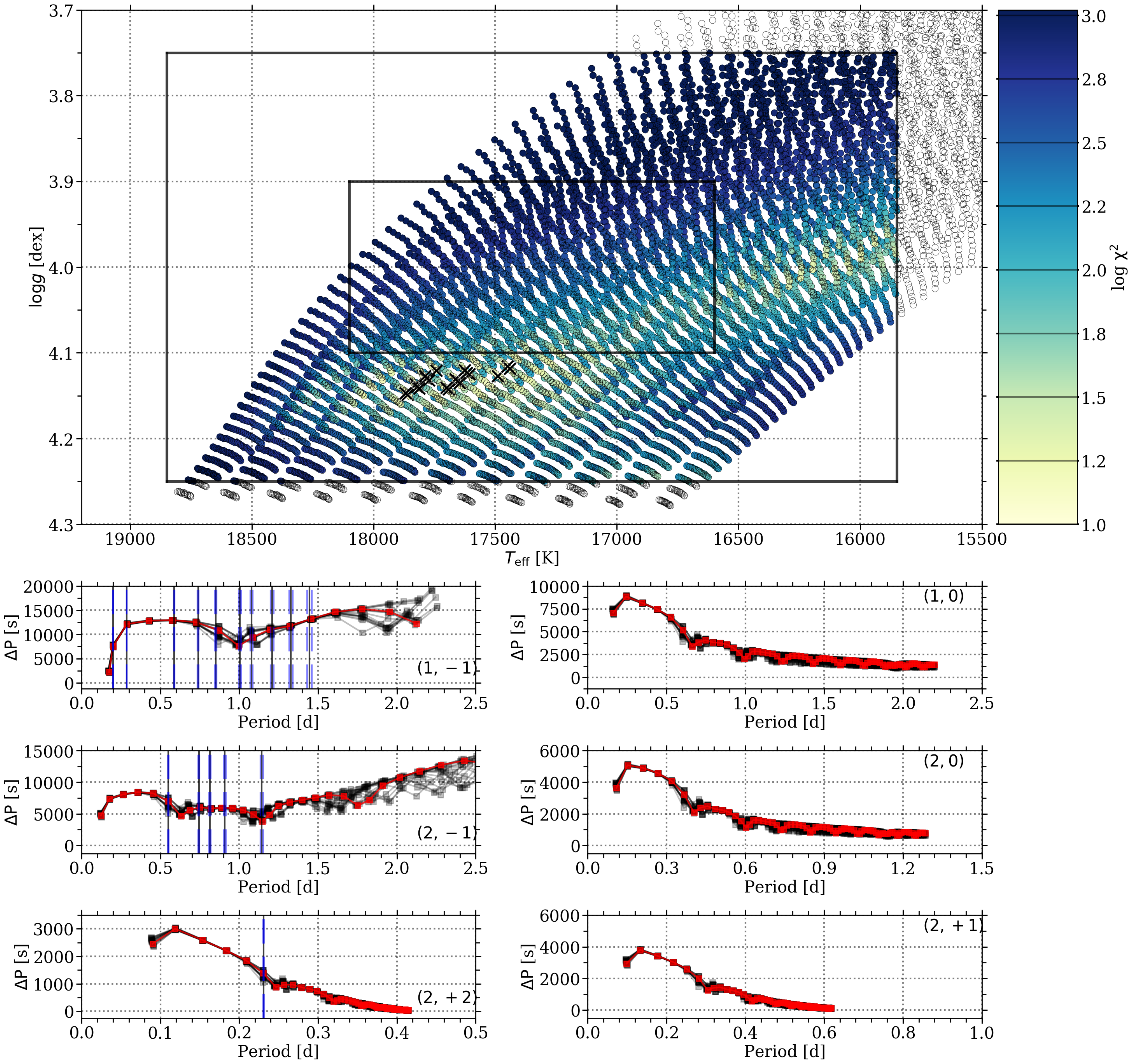}%
\caption{Summary plot for the overall forward modelling of HD\,43317 using the combined dipole mode hypothesis on the refined grid of MESA models.  The same colour scheme as Fig.\,\ref{fig:dipole_GYRE} is used.}
\label{fig:combined_GYRE}
\end{figure*}

At this stage, the theoretical frequencies of the best MESA models were able to explain eleven of the 28 observed frequencies of HD\,43317.  We studied whether yet unexplained detected frequencies could be associated to pulsation mode frequencies with a different geometry than $(1, -1)$ modes.  The frequency density argument made in Sect.\,\ref{sec:modelling_blind} and indicated in Fig.\,\ref{fig:GYRE_allmodes}, showed that it was impossible to resolve zonal and most $m > 0$ modes with the frequency resolution of the CoRoT data set.  Hence, we investigated whether theoretically predicted $(\ell, m) = (2, -1)$ modes could explain the additional detected frequencies.  Such modes have an angle of least cancellation at $45^\circ$ \citep{2010aste.book.....A}. This was close to the updated value for the inclination angle $i = 37 \pm 3\,^{\circ}$ derived in Sect.\,\ref{sec:discussion_params}, validating this assumption. We identified five matches between theoretically predicted $(2, -1)$ mode frequencies and the so far unused frequencies of HD\,43317, namely $f_{7}$, $f_{11}$, $f_{14}$, $f_{15}$ and $f_{21}$.  These were subsequently included in the hypothesis and the modelling was repeated.

\begin{figure*}[t]
\centering
\includegraphics[width=\textwidth, height = 0.5\textheight]{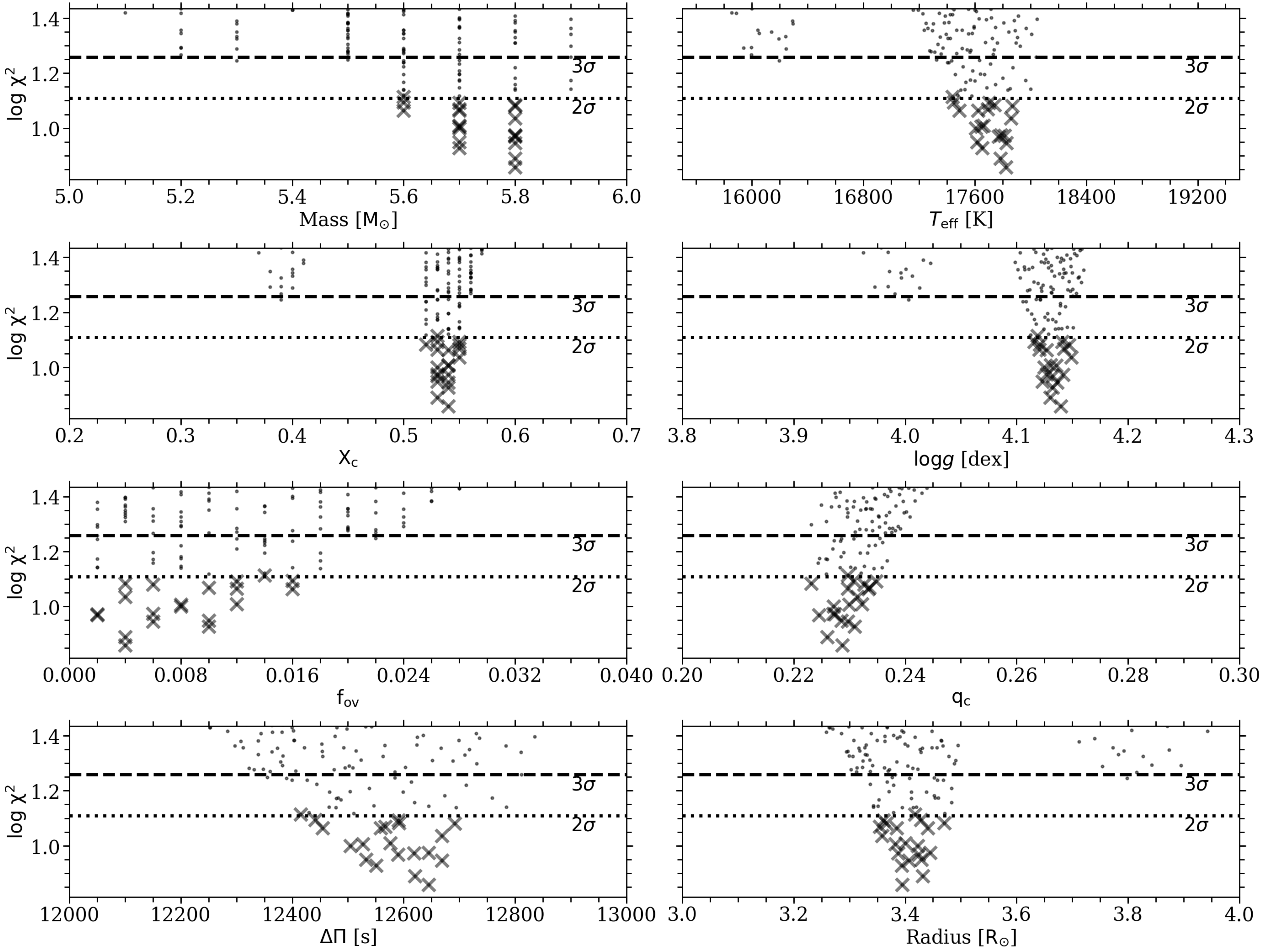}%
\caption{Distribution of the $\chi^2$ values, computed during the forward seismic modelling of HD\,43317 based on 16 of the 28 observed frequencies, using the refined grid of MESA models.  The $\chi^2$ values for the $2\sigma$ and $3\sigma$ confidence intervals are indicated by the dotted and dashed black line, respectively.  The best 20 models are indicated by the black crosses.  We refer to Table\,\ref{tab:bestmodels} for the description of the physical quantities.}
\label{fig:combined_chisquare}
\end{figure*}

The result of this final modelling under the hypothesis that only $(1, -1)$, $(2,-1)$ and $(2, +2)$ modes were observed in the CoRoT photometry of HD\,43317 is summarized in Fig.\,\ref{fig:combined_GYRE}.  The location of the 20 best fitted MESA models did not change by including the five additional $(2, -1)$ mode frequencies.  The three estimated physical parameters, as well as some other output variables, are listed in Table\,\ref{tab:bestmodels} and the $\chi^2$ distributions for these parameters are shown in Fig.\,\ref{fig:combined_chisquare}.  Table\,\ref{tab:bestmodel_frequency} compares the observed frequencies with those predicted by GYRE based on the MESA model with the lowest $\chi^2$. This model has parameters $M_{\star}{=}5.8$\,$\mathrm{M_{\odot}}$, $X_c{=}0.54$, and $f_{\rm ov}{=}0.004$.

\begin{table*}[t]
  \caption{Parameters of the 20 best models based on the identification of 16 out of the 28 detected frequencies of HD\,43317, ordered by the resulting $\chi^2$ value.}
\centering
\tabcolsep=6pt
\begin{tabular}{llllllllll}
\hline
\hline
$\chi^2$ & $M_{\star}$	 & $X_c$	 & $f_{\rm ov}$ & $R_{\star}$ & $\Delta \Pi$ & Age & $T_{\rm eff}$ & $\log g$ & $q_c$\\
		 & [$\mathrm{M_{\odot}}$] & & & [$\mathrm{R_{\odot}}$] & [s] & [Myr] & [K] & [dex] &\\
\hline
7.23		&	5.8	&	0.54	&	0.004	&	3.39	&	12645	&	28.4	&	17822	&	4.14	&	0.229	\\
7.76		&	5.8	&	0.53	&	0.004	&	3.43	&	12619	&	29.8	&	17782	&	4.13	&	0.226	\\
8.46		&	5.7	&	0.54	&	0.010	&	3.39	&	12550	&	31.9	&	17652	&	4.13	&	0.231	\\
8.82		&	5.8	&	0.54	&	0.006	&	3.41	&	12668	&	29.2	&	17826	&	4.14	&	0.230	\\
8.89		&	5.7	&	0.53	&	0.010	&	3.43	&	12532	&	33.3	&	17617	&	4.12	&	0.229	\\
9.31		&	5.8	&	0.53	&	0.002	&	3.42	&	12590	&	29.2	&	17772	&	4.13	&	0.225	\\
9.40		&	5.8	&	0.54	&	0.002	&	3.39	&	12618	&	27.8	&	17812	&	4.14	&	0.227	\\
9.44		&	5.8	&	0.53	&	0.006	&	3.44	&	12645	&	30.6	&	17788	&	4.13	&	0.227	\\
10.00		&	5.7	&	0.53	&	0.008	&	3.42	&	12504	&	32.7	&	17605	&	4.13	&	0.227	\\
10.15		&	5.7	&	0.54	&	0.008	&	3.38	&	12526	&	31.1	&	17646	&	4.14	&	0.230	\\
10.21		&	5.7	&	0.54	&	0.012	&	3.40	&	12575	&	32.6	&	17663	&	4.13	&	0.232	\\
10.87		&	5.8	&	0.55	&	0.004	&	3.36	&	12668	&	27.0	&	17859	&	4.15	&	0.231	\\
11.58		&	5.6	&	0.54	&	0.016	&	3.38	&	12454	&	35.3	&	17489	&	4.13	&	0.233	\\
11.61		&	5.7	&	0.53	&	0.012	&	3.44	&	12558	&	34.2	&	17624	&	4.12	&	0.230	\\
11.72		&	5.7	&	0.55	&	0.010	&	3.35	&	12567	&	30.3	&	17690	&	4.14	&	0.234	\\
12.05		&	5.8	&	0.55	&	0.006	&	3.37	&	12691	&	27.6	&	17868	&	4.15	&	0.233	\\
12.11		&	5.8	&	0.52	&	0.004	&	3.47	&	12591	&	31.2	&	17742	&	4.12	&	0.223	\\
12.37		&	5.7	&	0.55	&	0.012	&	3.36	&	12590	&	30.9	&	17701	&	4.14	&	0.235	\\
12.41		&	5.6	&	0.53	&	0.016	&	3.43	&	12441	&	37.1	&	17449	&	4.12	&	0.231	\\
13.00		&	5.6	&	0.53	&	0.014	&	3.42	&	12415	&	36.3	&	17440	&	4.12	&	0.230	\\
\hline
\end{tabular}
\label{tab:bestmodels}
\tablefoot{For each MESA model, the stellar mass, $M_{\star}$, central hydrogen mass fraction, $X_c$, convective core overshooting parameter, $f_{\rm ov}$, stellar radius, $R_{\star}$, asymptotic period spacing, $\Delta \Pi$, age, effective temperature, $T_{\rm eff}$, surface gravity, $\log g$, and fractional mass of the convective core, $q_c$, are given.  Degeneracies among the stellar parameters occur when only considered the value for the asymptotic period spacing.}
\end{table*}

\begin{table*}[t]
\caption{Comparison between the observed frequencies, $f_{\rm obs}$, and the GYRE frequencies of the MESA model resulting in the best description of the CoRoT data of HD\,43317, $f_{n_g, l, m}$.  We compared the frequency shift due to rotation according to the traditional approximation with the frequency shift due to an internal (purely) poloidal magnetic field.}
\centering
\tabcolsep=6pt
\begin{tabular}{llrlrl|lll|llcc}
\hline
\hline
		& $f_{\rm obs}$	& $n_g$ & $\ell$ & $m$ & $f_{n_g, l, m}$	& $f_{n_g, l, 0}$ & $f_{n_g, l, 0}$ & $f_{\rm shift}$	& $f_{\rm shift}$ & $f_{\rm shift}$ & $S_c$	& $\mathcal{I}$\\
		&			&		&		&		& model			& rot. 			& no rot.	 	   & rot.				& 26.1\,kG			& 82.4\,kG		\\
		& [\d]		&		&		&		& [\d]			& [\d]			& [\d]				&[\d]				& [\d]				& [\d]			& [$\mathrm{G^{-2}}$]&	\\
\hline
$f_{4}	$&	0.6916	&	-11	&	1	&	-1	&	0.6867	&	1.1927	&	0.8187	&	0.3740	&	0.0210	&	0.2092	&$	3.7661	\cdot10^{-11}	$&$	1.5487	\cdot10^{7}	$\\
$f_{5}	$&	0.7529	&	-10	&	1	&	-1	&	0.7573	&	1.2595	&	0.8979	&	0.3617	&	0.0159	&	0.1592	&$	2.6131	\cdot10^{-11}	$&$	1.2925	\cdot10^{7}	$\\
$f_{6}	$&	0.8278	&	-9	&	1	&	-1	&	0.8381	&	1.3388	&	0.9923	&	0.3465	&	0.0125	&	0.1250	&$	1.8565	\cdot10^{-11}	$&$	1.1215	\cdot10^{7}	$\\
$f_{7}	$&	0.8752	&	-15	&	2	&	-1	&	0.8720	&	1.6623	&	1.0462	&	0.6161	&	0.0110	&	0.1101	&$	1.5515	\cdot10^{-11}	$&$	4.3758	\cdot10^{7}	$\\
$f_{8}	$&	0.9279	&	-8	&	1	&	-1	&	0.9222	&	1.4236	&	1.0933	&	0.3303	&	0.0104	&	0.1041	&$	1.4035	\cdot10^{-11}	$&$	1.0293	\cdot10^{7}	$\\
$f_{9}	$&	0.9954	&	-7	&	1	&	-1	&	1.0037	&	1.5086	&	1.1943	&	0.3143	&	0.0097	&	0.0972	&$	1.1994	\cdot10^{-11}	$&$	1.0496	\cdot10^{7}	$\\
$f_{11}	$&	1.1004	&	-11	&	2	&	-1	&	1.1268	&	1.9545	&	1.4108	&	0.5438	&	0.0029	&	0.0288	&$	3.0120	\cdot10^{-12}	$&$	1.5448	\cdot10^{7}	$\\
$f_{13}	$&	1.1754	&	-6	&	1	&	-1	&	1.1483	&	1.6594	&	1.3708	&	0.2887	&	0.0086	&	0.0861	&$	9.2633	\cdot10^{-12}	$&$	1.0679	\cdot10^{7}	$\\
$f_{14}	$&	1.2280	&	-10	&	2	&	-1	&	1.2198	&	2.0621	&	1.5454	&	0.5167	&	0.0022	&	0.0220	&$	2.0968	\cdot10^{-12}	$&$	1.2904	\cdot10^{7}	$\\
$f_{15}	$&	1.3424	&	-9	&	2	&	-1	&	1.3337	&	2.1920	&	1.7058	&	0.4861	&	0.0017	&	0.0173	&$	1.4932	\cdot10^{-12}	$&$	1.1195	\cdot10^{7}	$\\
$f_{16}	$&	1.3529	&	-5	&	1	&	-1	&	1.3775	&	1.8992	&	1.6455	&	0.2537	&	0.0063	&	0.0626	&$	5.6119	\cdot10^{-12}	$&$	9.3225	\cdot10^{6}	$\\
$f_{20}	$&	1.7045	&	-4	&	1	&	-1	&	1.7358	&	2.2719	&	2.0601	&	0.2118	&	0.0045	&	0.0451	&$	3.2246	\cdot10^{-12}	$&$	8.3961	\cdot10^{6}	$\\
$f_{21}	$&	1.8156	&	-6	&	2	&	-1	&	1.8191	&	2.7238	&	2.3408	&	0.3829	&	0.0012	&	0.0121	&$	7.5960	\cdot10^{-13}	$&$	1.0725	\cdot10^{7}	$\\
$f_{29}	$&	3.4958	&	-2	&	1	&	-1	&	3.4811	&	4.1058	&	3.9638	&	0.1420	&	0.0022	&	0.0221	&$	8.2084	\cdot10^{-13}	$&$	7.9125	\cdot10^{6}	$\\
$f_{31}	$&	4.3311	&	-6	&	2	&	2	&	4.3408	&	2.7238	&	2.3408	&	0.3829	&	0.0048	&	0.0483	&$	3.0384	\cdot10^{-12}	$&$	1.0725	\cdot10^{7}	$\\
$f_{32}	$&	5.0047	&	-1	&	1	&	-1	&	4.9948	&	6.1855	&	6.1084	&	0.0771	&	0.0015	&	0.0150	&$	3.6143	\cdot10^{-13}	$&$	8.2739	\cdot10^{6}	$\\
\hline
\end{tabular}
\label{tab:bestmodel_frequency}
\tablefoot{For each observed frequency, we provide the best model frequency for the indicated mode geometry, as well as the corresponding zonal pulsation mode frequencies $f_{n_g, l, 0}$ computed with and without the Coriolis force in the traditional approximation.  The frequency shift due to the Lorentz force was computed following the approach of \citet{2005A+A...444L..29H}, for the lower and upper limit of an internal magnetic field at the convective core boundary, by extrapolating the surface field inward.  We also included the magnetic splitting coefficient $S_c$ and the quantity $\mathcal{I}$.}
\end{table*}

\section{Discussion}
\label{sec:discussion}

\subsection{Seismic estimation of the stellar parameters}
\label{sec:discussion_params}
\begin{figure}
\centering
\includegraphics[width=0.45\textwidth, height = 0.33\textheight]{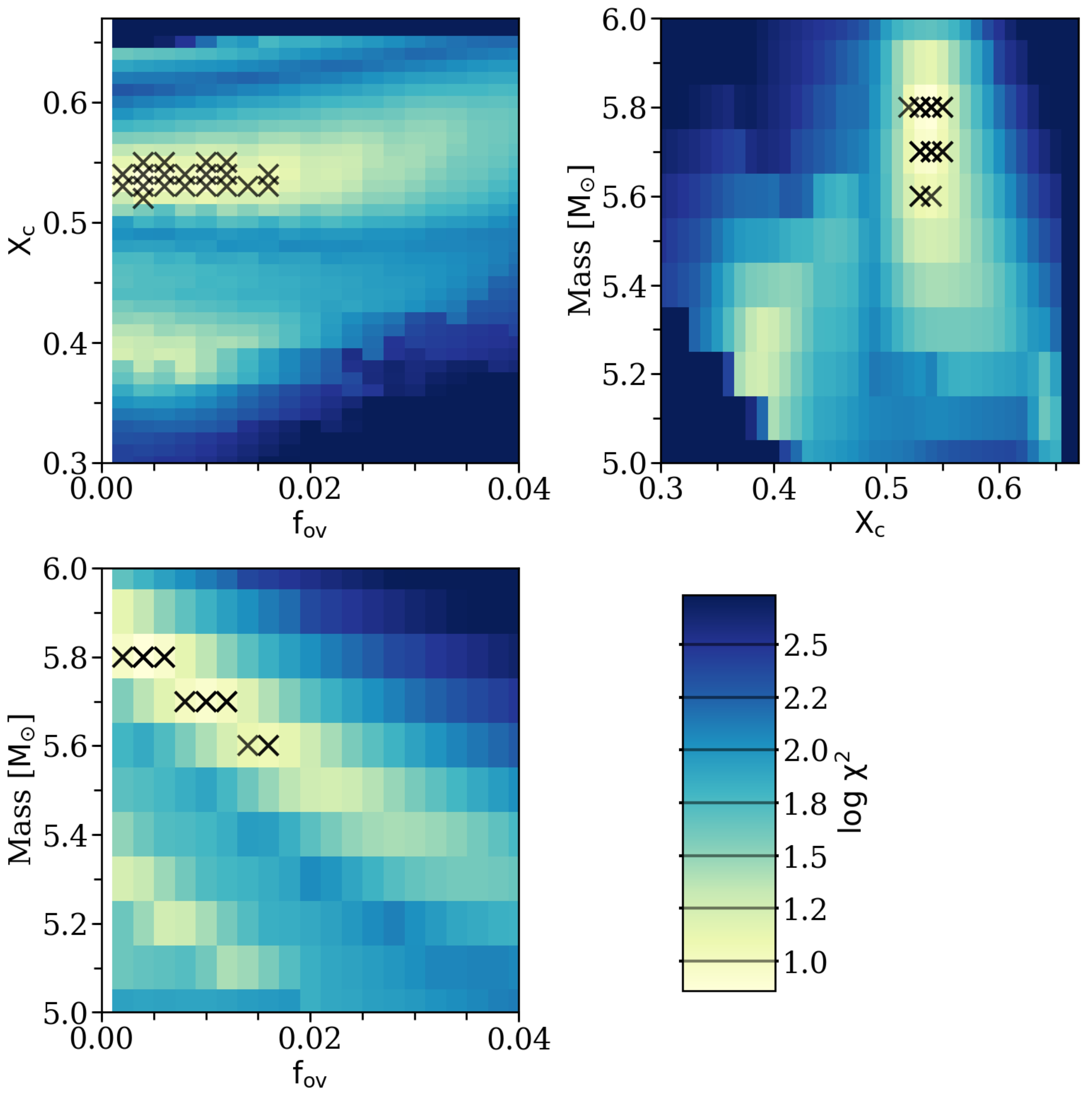}%
\caption{Colour maps indicating the pairwise (marginalized) correlation between the parameters  in the (fine) grid of MESA models.  \textit{Top left}: correlation between $X_c$ and $f_{\rm ov}$ after marginalizing over $M_{\star}$.  \textit{Top right}: correlation between $M_{\star}$ and $X_c$ after marginalizing over $f_{\rm ov}$. \textit{Bottom left}: correlation between $M_{\star}$ and $f_{\rm ov}$ after marginalizing over $X_c$, indicating the strong correlation caused by models with the same $\Delta \Pi$ value.  The same colour scale was used for all panels. The 20 best MESA models are indicated by the black crosses.}
\label{fig:combined_marg}
\end{figure}

Instead of using the discrete step size of the grid as the confidence intervals for $M_{\star}$, $X_c$, and $f_{\rm ov}$ as deduced from the forward modelling, we used the properties of the $\chi^2$ statistics.  We computed the upper limit on the $\chi^2$ value for a $2\sigma$ confidence interval (i.e., $95.4$\,\% confidence interval) using Eq.\,(\ref{eq:upperlimit}).  We included 16 frequencies in our final fitting process, resulting in 12 degrees of freedom, which led to $\chi^2_{2\sigma}{=}12.85$.  This corresponds to the $\chi^2$ value of the best 19 MESA models (as shown in Fig.\,\ref{fig:combined_chisquare}), supporting the visual inspection of the best models during the procedure.  The corresponding confidence intervals on the parameters were $M_{\star}{=} 5.8^{+0.1}_{-0.2}$\,$\mathrm{M_{\odot}}$, $X_c{=}0.54^{+0.01}_{-0.02}$, and $f_{\rm ov}{=}0.004^{+0.014}_{-0.002}$.  The skewed confidence intervals for $M_{\star}$ and $f_{\rm ov}$ resulted from a strong correlation between these parameters, as demonstrated in Fig.\,\ref{fig:combined_marg}.  We note that our grid of MESA models was limited from $f_{\rm ov}{=}0.002$ to 0.040, hence we could not exclude that $f_{\rm ov}$ had an even lower value than 0.002, in particular $f_{\rm ov}=0.0$.  The MESA models within this $\chi^2$ valley had comparable values for the asymptotic period spacing pattern (see Table\,\ref{tab:bestmodels}) defined as:
\begin{equation}
\Delta \Pi = \frac{\pi}{\int \frac{N}{r} \mathrm{d}r} \,\mathrm{,}
\label{eq:periodspacing}
\end{equation}
\noindent where $N$ is the Brunt-V\"{a}is\"{a}l\"{a} frequency, the integral is over the g-mode cavity, and $\Delta \Pi$ is given in seconds (omitting the factor $2\pi$ in the definition).  Comparable $\Delta \Pi = 12650^{+50}_{-250}$\,s values were caused by a similar value for the integral, following from the relation between the overall stellar mass and the mass inside the core and convective core overshooting region.  We were unable to lift the degeneracy caused by this correlation without many more observed and well defined pulsation modes that had a different probing power in the near-core region.  Additional observed trapped modes would be particularly helpful to achieve this, as they are most sensitive to the near-core layers.  Such modes manifest themselves in the regions of the local minima in the period spacing patterns as observed with the \textit{Kepler\/} satellite for numerous g-mode pulsators, but {\it Kepler\/} data have a ten times better frequency resolution than our CoRoT data.

\citet{2012A+A...542A..55P} used the few dominant modes to estimate $\Delta \Pi$, without modelling the individual frequencies and while ignoring the rotation of the star. It turned out that we obtained a value about twice as high, pointing out the difficulty of deducing an appropriate $\Delta \Pi$ value without good knowledge of the rotation frequency of the star. We emphasize that the high-precision $f_{\rm rot}$ value of HD\,43317 determined from the magnetic modelling by \citet{2017A+A...605A.104B} was an essential ingredient for the successful seismic modelling of the star as presented here.

To investigate the pairwise correlation between the parameters of the grid of MESA models, we employed the marginalization technique to reduce the dimensionality of the $\chi^2$ landscape.  This technique took the minimum $\chi^2$ value of the fit along the third axis of the grid, providing easily interpretable correlation maps (given in Fig.\,\ref{fig:combined_marg}).  These correlation maps indicated that the central hydrogen fraction, $X_c$, was well constrained, but these maps also illustrated the tight correlation between $M_{\star}$ and $f_{\rm ov}$.  These correlation maps also illustrated the existence of the (non-significant) secondary solution of MESA models, which was already apparent from the $\chi^2$ landscape in the Kiel diagram (see Fig.\,\ref{fig:combined_GYRE}) and from the $\chi^2$ distribution of several parameters in Fig.\,\ref{fig:combined_chisquare}.  The bottom left panel of Fig.\,\ref{fig:combined_chisquare} demonstrated that these models had a comparable value for $\Delta \Pi $ to the global solution, thus explaining the reason why this secondary solution was (partly) retrieved.

The correlation between $f_{\rm ov}$ and $M_{\star}$ and the limited frequency resolution of the CoRoT data led to a relatively large confidence interval on the derived $f_{\rm ov}$ for HD\,43317.  The low value of the best model agreed well with the predictions from theory and simulations for hot stars hosting a detectable large-scale magnetic field and with the only other observational result \citep[i.e., V2052\,Oph,][]{2012MNRAS.427..483B}.  The $2\sigma$ confidence interval on $f_{\rm ov}$ was also compatible with the seismic estimate of this parameter for the non-magnetic B-type star KIC\,10526294 \citep[][see their Fig.\,B.1]{2015A+A...580A..27M}.  However, KIC\,10526294 is a slow rotator with a rotation period of about 190\,d, while HD\,43317 is a rapid rotator.  The result for $f_{\rm ov}$ from the seismic modelling of the B-type star KIC\,7760680 \citep[][their Fig.\,6]{2016ApJ...823..130M} with $P_{\rm rot} = 2.08 \pm 0.04$\,d differs more than $2\sigma$ from the $f_{\rm ov}$-value for HD\,43317.  Thus, the forward seismic modelling of HD\,43317 is compatible with the suppression of near-core mixing due to a large-scale magnetic field, but we could not exclude that other physical processes are involved in the limited overshooting, such as the near-core rotation.

By construction of the best models, the seismic estimates for $T_{\rm eff}$ and $\log g$ agreed well with the spectroscopic values by \citet{2012A+A...542A..55P}.  The age estimate for the best model (28.4\,Myr, see Table\,\ref{tab:bestmodels}) is compatible with a literature value obtained from isochrone fitting to the spectroscopic parameters \citep[i.e., $26\pm6$\,Myr;][]{2011MNRAS.410..190T}.  Both results rely on stellar models that were computed with independent codes.

Further, we redetermined the inclination angle $i$ using the radius of the best fitted MESA model (i.e., $R = 3.39$\,$\mathrm{R_{\odot}}$), the value for $v \sin i = 115 \pm 9$\,km\,s$^{-1}$ \citep{2012A+A...542A..55P} and the rotation period $P_{\rm rot} = 0.897673(4)$\,d \citep{2017A+A...605A.104B}.  This resulted in an inclination angle $i = 37 \pm 3$\,$^{\circ}$, where the largest uncertainty came from the estimate of $v \sin i$.  This a posteriori refinement of the inclination angle is compatible with the mode visibility of the dipole and quadrupole mode interpretation.  With this inclination angle, we derived an updated value for the obliquity angle $\beta = 81 \pm 6$\,$^{\circ}$, employing the measurements for their longitudinal magnetic field of \citet{2017A+A...605A.104B} for the LSD profiles with their complete line mask.  Using these angles, the longitudinal magnetic field measurements of \citet{2017A+A...605A.104B}, the equation for the dipolar magnetic field strength of \citet{1950ApJ...112..222S} and a linear limb-darking coefficient of $u=0.3$ (appropriate for a B3.5V star, see e.g., \citealt{2000A+A...363.1081C}), we deduced that the dipolar magnetic field of HD\,43317 had a strength $B_{\rm dip} = 1312 \pm 332$\,G.  These values were all consistent with the ranges obtained by \citet{2017A+A...605A.104B}, which were based on the range of acceptable inclination angles from \citet{2012A+A...542A..55P}.

\subsection{Dependences on the mode identification assumptions}
During the forward seismic modelling, we were explicit on the assumptions about the mode geometry, and at which stage a given observed frequency entered the modelling scheme.  Several of these frequencies did not have a unique mode identification (even for the best fitted MESA model).  As an example, we discuss $f_{15}$, which we assumed to be a $(\ell,m){=}(2,-1)$ mode.  However, the zonal mode frequency $f_{n_g, \ell, 0}$ of $f_6$ also matched closely with $f_{15}$.  Such a degeneracy occurred for several cases, but never in a systematic way.  Moreover, these degeneracies did not significantly alter the confidence intervals of the parameters based on the selected MESA models as their corresponding GYRE frequencies had a good match with the observations.  We considered it more sensible to use interrupted series of $(1, -1)$ and $(2, -1)$ modes (together with one $(2,+2)$ mode confirmed by spectroscopy) during the forward seismic modelling of HD\,43317 than injecting zonal modes that did not belong to any series in radial order or rotationally split multiplets.   Finally, we emphasize that the assumption of the initial four $(1, -1)$ modes with the spectroscopic identification of $f_{31}$ and the enforced $2\sigma$ spectroscopic box sufficiently constrained the possible MESA models for HD\,43317, while still complying with forward modelling results of {\it Kepler\/} B-type g-mode pulsators.

\subsection{Assessment of the theoretical pulsation mode frequencies}
In our computations of the mode frequencies, we made two important assumptions.  The first one was uniform rotation, as the GYRE computations used the traditional approximation under this assumption.  Following the results in \citet{2017ApJ...847L...7A} and in Van Reeth et al. (submitted) for tens of g-mode pulsators, this was a valid assumption.  Furthermore, uniform rotation in the radiative layer was theoretically predicted for stars with a stable large-scale magnetic field \citep[e.g.,][]{1937MNRAS..97..458F, 1992MNRAS.257..593M, 1999A+A...349..189S, 2005A+A...440..653M, 2011IAUS..272...14Z}.

The second assumption was that we ignored the effect of the magnetic field on the g-mode frequencies because we did not have observational information on the interior properties of the large-scale magnetic field.  Below we attempted to assess the consequences of these assumptions after we obtained a good model representing the stellar structure of the pulsating magnetic star HD\,43317.

\subsubsection{Frequency shifts of zonal modes due to the Coriolis force}
For HD\,43317, the independent measurement of the rotation frequency (at the stellar surface) was a necessary pre-requisite to be able to perform seismic modelling.  Indeed, the well-known estimation procedure for $f_{\rm rot}$ from the period spacing patterns of g modes applied to {\it Kepler\/} data as in \citet{2016ApJ...823..130M}, \citet{2016A+A...593A.120V} or \citet{2017MNRAS.465.2294O} was not possible for our CoRoT target.  Prior knowledge of the rotation frequency allowed us to apply the traditional approximation for rotation, which ignores the latitudinal component of the rotation vector. This approximation is appropriate for g modes in B- and F-type stars thanks to their large horizontal displacements, as long as they do not rotate too close to their critical rotation velocity \citep{2017MNRAS.465.2294O}.  However, forward modelling of such pulsators was so far also still done without taking into account the Coriolis force for slow rotators \citep[cf.,][]{2015A+A...580A..27M,2016A+A...592A.116S}.

Here, we wish to compare the effects of ignoring the Coriolis and Lorentz forces for the resulting theoretical predictions of the g-mode frequencies, taking the case of HD\,43317 as the only concrete example of a magnetic  g-mode pulsator so far. Such comparison is most easily done for zonal modes (i.e., $m{=}0$) of stellar models, as these are not subject to transformation effects between the co-rotating and inertial reference frames.

Unlike a first-order perturbative approach for the effects of the Coriolis force \citep[e.g.,][]{1951ApJ...114..373L}, the traditional approximation results in frequency shifts for zonal pulsation modes.  We computed the frequency differences for the zonal mode frequencies $f_{n_g, \ell, 0}$ of the identified modes in the best MESA model with $M_{\star}{=} 5.8$\,$\mathrm{M_{\odot}}$, $X_c{=}0.54$, and $f_{\rm ov}{=}0.004$, when ignoring and while taking into account the Coriolis force.  The results are listed in Table\,\ref{tab:bestmodel_frequency}.  As expected, the frequency shift due to the Coriolis force is large for the rotation rate of HD\,43317.  It increases with increasing radial order $n_g$ and with mode degree $\ell$.  For the pulsation modes with the highest $n_g$ values, the frequency-shifted value easily exceeded 25\,\% of the non-rotating value $f_{n_g, \ell, 0}$, clearly illustrating the need to account for the stellar rotation during the forward seismic modelling.

\subsubsection{Frequency shifts of zonal modes due to the Lorentz force}
To compute the shift in the pulsation mode frequency caused by an internal magnetic field, we followed the perturbative approach of \citet{2005A+A...444L..29H}, since it was one of the few available formalisms applicable to g modes.  It assumes that the internal magnetic field corresponds to a poloidal axisymmetric field.  While this is a limitation, there is currently no better prescription to apply.

Theoretical studies and numerical simulations showed that any extension of a large-scale magnetic field measured at the stellar surface towards the stellar interior needs to have both a toroidal and poloidal component of about equal strength to be stable over long time scales \citep[e.g.,][]{1980MNRAS.191..151T, 2006A+A...450.1077B, 2010A+A...517A..58D,   2010ApJ...724L..34D}.  However, \citet{2005A+A...444L..29H} argued that the toroidal component of the internal magnetic field would lead to a lower frequency shift for high-radial order g modes than the poloidal component.  Hence, we adopted their formulation here.  The resulting frequency shift can be expressed as
\begin{equation}
\frac{\delta \omega}{\omega} = \frac{1}{2}\left(\frac{\omega_{\rm A}}{\omega}\right)^2~C_{l,m}~\mathcal{I} = S_{\rm c}~B^2_0 \,\mathrm{,}
\label{eq:magnetic_shift}
\end{equation}
\noindent where
\begin{equation}
\omega_{\rm A}=\frac{B_0}{\sqrt{4\pi\rho_c}}\frac{1}{R_{\star}}
\label{eq:alfven_frequency}
\end{equation} 
\noindent is the Alfv\'en frequency for an internal magnetic field with strength $B_0$.  This expression led to the magnetic splitting coefficient $S_{\rm c}$ given as
\begin{equation}
S_{\rm c} = \frac{C_{\ell, m} \mathcal{I}}{8 \pi \omega^2 \rho_{\rm c} R_{\star}^2} \,\mathrm{,}
\label{eq:magnetic_coefficient}
\end{equation}
\noindent where $\rho_{\rm c}$ is the central mass density, $R_{\star}$ is the stellar radius, $\omega$ is the cyclic frequency (in rad\,s$^{-1}$) corresponding to the angular pulsation mode frequency $f_{n_g, \ell, m}$ and the Ledoux coefficients $C_{\ell, m}$ have been introduced (see \citealt{1957AnAp...20..185L} and Eq.\,(8) and (9) of \citealt{2005A+A...444L..29H}).  They describe the horizontal overlap between the g-mode displacement, assumed to be predominantly horizontal, with the dipolar magnetic field.  Finally, $\mathcal{I}$ is defined as
\begin{equation}
\mathcal{I} = \frac{\int \left| \frac{2}{x} \frac{\mathrm{d}}{\mathrm{d}x} \left(x\,b(x)\,\xi_{\rm h}(x)\right)  \right|^2 x^2 \,\mathrm{d}x}{\int \left| \xi_{\rm h}(x)\right|^2 \frac{\rho(x)}{\rho_{\rm c}} x^2 \,\mathrm{d}x}\,\mathrm{,}
\label{eq:magnetic_I}
\end{equation}
\noindent with $x=r/R_{\star}$ the radial coordinate, $\xi_{\rm h}(x)$ the horizontal displacement for the pulsation mode with frequency $f_{n_g, \ell, m}$, $\rho(x)$ the internal density profile, and $b(x)$ the profile of the magnetic field as a function of the radial coordinate.   We followed the definition of \citet{2005A+A...444L..29H} and assumed $b(x) = \left(x/x_{\rm c}\right)^{-3}$, with $x_{\rm c}$ the radial coordinate of the outer edge of the convective core.  For our best model of HD\,43317 ($M_{\star}{=}5.8$\,$\mathrm{M_{\odot}}$, $X_c{=}0.54$, and $f_{\rm ov}{=}0.004$), the MESA model delivered $x_{\rm c} = 0.168$.

\begin{figure}
\centering
\includegraphics[width=0.45\textwidth, height = 0.33\textheight]{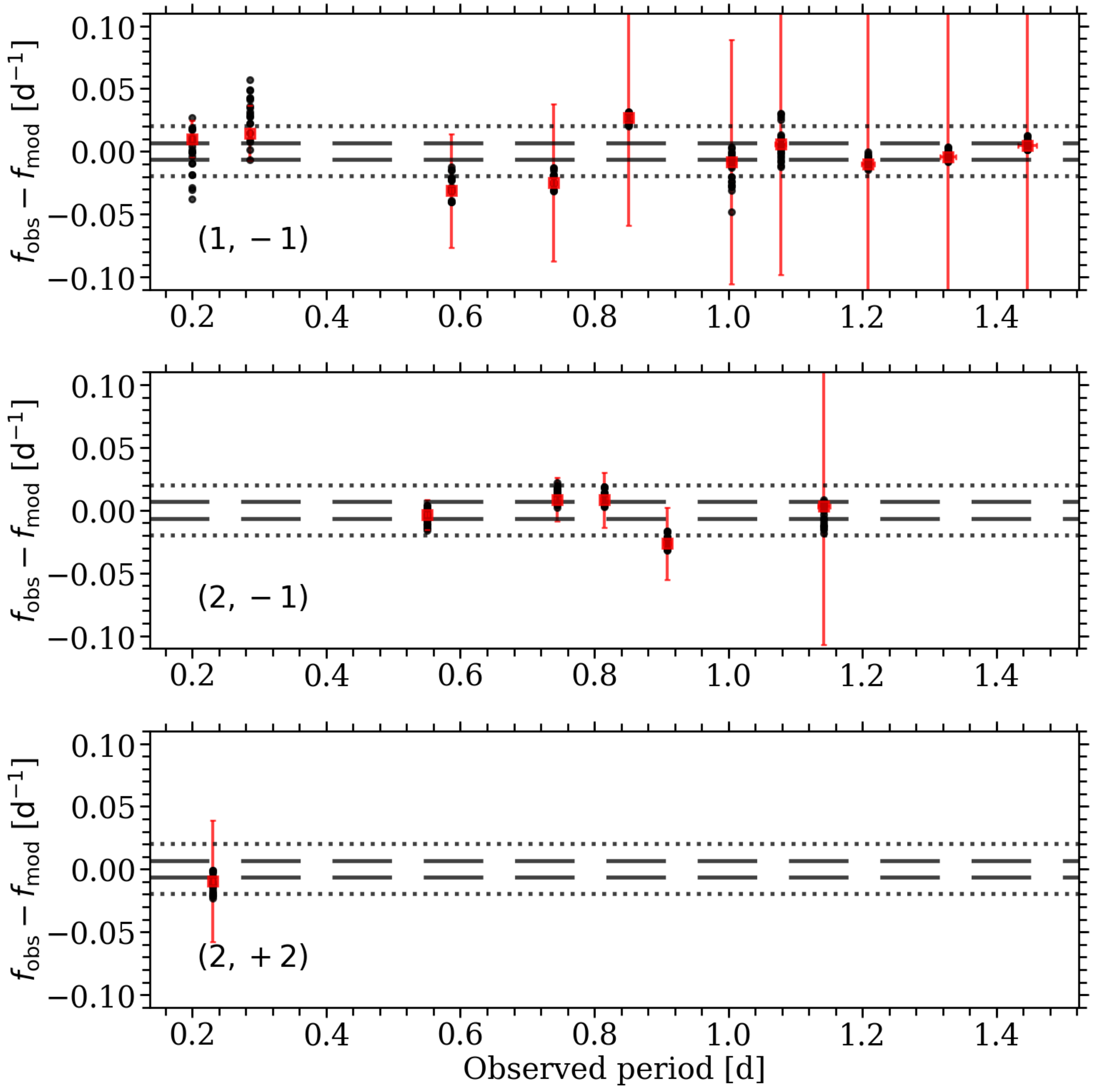}%
\caption{Difference in frequency between the observed pulsation mode frequencies and the GYRE frequencies of the 20 best MESA models, ordered according to the mode geometry.  The red squares indicate the difference between the observations and the best model description.  The red vertical errorbars indicate the frequency shift caused by an internal poloidal magnetic field of 82.4\,kG and the red horizontal error bars represent the Rayleigh frequency resolution of the CoRoT light curve.  Both were often similar to the symbol size, while the magnetic frequency shift for the highest period dipole modes was larger than the indicated panel (see also Table\,\ref{tab:bestmodel_frequency}).  The black dashed and dotted lines show the Rayleigh frequency resolution and $3\delta f_{\rm ray}$, respectively.}
\label{fig:freqdiff}
\end{figure}

We estimated the strength of the frozen-in large-scale magnetic field of HD\,43317 at $x_{\rm c}$ following the results provided by the simulations of \citet[][we refer the reader to his Fig.\,8]{2008MNRAS.386.1947B}.  Depending on the age of the star, the internal magnetic field was 26.6 to 50.1 times as strong as at the stellar surface.  Using our new estimate of the strength of the dipolar magnetic field at the surface of HD\,43317, $B_{\rm dip} = 1312 \pm 332$\,G, we got a near-core magnetic field strength in the range $B_0{=}26.1 - 82.4$\,kG.  We computed the frequency shift using Eq.\,\ref{eq:magnetic_shift} for these two limiting values of $B_0$, the model frequencies $f_{n_g, \ell, 0}$ of the non-rotating case (see Table\,\ref{tab:bestmodel_frequency}) and the difference in Ledoux constants $\Delta C_{\ell, m}$ values to account for the mode geometry \citep[similarly to ][]{2005A+A...444L..29H}.  The values for the obtained frequency shifts due to the Lorentz force are given in Table\,\ref{tab:bestmodel_frequency} and are compared with the differences between the observed and the 20 best sets of model frequencies in Fig.\,\ref{fig:freqdiff}.

As expected from Eq.\,(\ref{eq:magnetic_shift}), we found that the frequency shifts depended on the strength of $B_0$.  Moreover, they increased with increasing radial order, since $\omega$ decreased and $\mathcal{I}$ increased.  We found that the frequency shift was largest for the observed $(1, -1)$ modes, since the difference in the Ledoux constants $\Delta C_{\ell, m}$ were largest for such modes.  All these results were compatible with those of \citet{2005A+A...444L..29H}.

The upper limit on the frequency shift due to the Lorentz force was almost always an order of magnitude smaller than that caused by the Coriolis force.  Hence, correctly accounting for the (uniform) rotation rate remained not only necessary, but was also more important during the forward modelling than accounting for a possible internal magnetic field that resulted from extrapolation of the surface field strength when dealing with field amplitudes as the one measured for HD\,43317.  This can also be further understood when revisiting Eq.\,(\ref{eq:magnetic_shift}).  First, the ratio of the Alfv\'en frequency and the pulsation mode frequencies was always very small (typically of the order of $10^{-9}$) permitting us to adopt a perturbative treatment of the effect of the Lorentz force on the pulsation modes.  Most of the observed modes during the modelling were sub-inertial gravito-inertial modes (i.e. $f<2 f_{\rm rot}$).  Therefore, they would be confined to an equatorial belt and thus propagate above a critical colatitude $\theta_c=\arccos(f/(2f_{\rm rot}))$ \citep[see e.g.,][]{2003MNRAS.343..125T}.  In the formalism by \citet{2005A+A...444L..29H}, the non-rotating approximation was assumed and modes were thus expanded as spherical harmonics corresponding to modes propagating in the whole sphere.  In this framework, the coefficient $C_{l,m}$ evaluated the horizontal overlap of the oscillation mode with the magnetic field with positive integrals \citep[we refer the reader to Eq.\,(7) in][]{2005A+A...444L..29H}.  As such, the determined frequency shifts due to the magnetic field should be larger in the case of the non-rotating approximation compared to equatorially trapped sub-inertial gravito-inertial modes.  The modification of the radial term from the non-rotating case to the case of a sub-inertial case is more difficult to infer. However, if $\omega_{\rm A}\ll 2\Omega$, it would not change the fact that the perturbative approach for the Lorentz force can be used.

In the case of super-inertial gravito-inertial modes (i.e., $f>2f_{\rm rot}$), the situation would be much simpler since waves propagate in the whole sphere as in the non-rotating case and the eigenmodes would be less modified by the Coriolis acceleration \citep[e.g.,][]{1997ApJ...491..839L}.  Therefore, the evaluation of the frequency shift provided by \citet{2005A+A...444L..29H} using the non-rotating assumption could be considered as a first reasonable step to evaluate the frequency shift induced by the Lorentz force.  The next step would be to generalise their work using the traditional approximation to take into account the effect of the Coriolis acceleration on the oscillation modes, in particular in the sub-inertial regime. This would require a dedicated theoretical study and was far beyond the scope of the current paper.

We concluded, as an a posteriori check, that the working principle of our seismic modelling was acceptable for this star.  Only for the higher-order dipole modes did the upper limit of the magnetic frequency shift become comparable with the frequency shift caused by the Coriolis force.  Yet, only a few such pulsation mode frequencies were identified for HD\,43317.  In this case a non-perturbative approach, as the one derived in \citet{2011A+A...526A..65M}, should be adopted.

Comparing the magnetic frequency shift with the differences between the observed and model frequencies (see Fig.\,\ref{fig:freqdiff}) led to the conclusion that the current description of the frequency shifts due to the Lorentz force failed to improve the differences. This was not surprising, given the simplification of the perturbative approach by \citet{2005A+A...444L..29H}, its assumptions on the geometry of the large-scale magnetic field, and because it neglects the Coriolis acceleration.  Future inclusion of the frequency shifts due to a magnetic field under the traditional approximation would be valuable for proper seismic modelling of magnetic g-mode pulsators and to estimate the upper limit of the interior magnetic field strength for stars without measurable surface field.

\section{Summary and conclusions}
\label{sec:conclusion}
We performed forward seismic modelling of the only known magnetic B-type star exhibiting independent g-mode pulsation frequencies.  The modelling was based on a grid of non-rotating, non-magnetic 1D stellar evolution models (computed with MESA) coupled to the adiabatic module of the pulsation code GYRE, while accounting for the uniform rotation and using the traditional approximation. This procedure allowed us to explain 16 of the 28 independent frequencies determined from the ${\sim}150$\,d CoRoT light curve.  We identified these 16 pulsation mode frequencies as ten $(1,-1)$ modes, one $(2,2)$ mode, and five $(2, -1)$ modes.  With this, the star revealed two overlapping period spacing series.  Other than $f_{31}$, zonal and prograde pulsation modes had to be excluded during the forward modelling, because the frequency resolution of the CoRoT light curve did not permit us to identify the theoretical counterparts of detected pulsation mode frequencies.

Some degeneracy on the mode geometry remained for a few of the used frequencies but this did not affect the three estimated stellar parameters, given their confidence intervals deduced from the best   models.  Most of the high frequencies in the CoRoT data were explained as rotationally shifted g modes. Hence, the interpretation by \citet{2012A+A...542A..55P} of having detected isolated p modes in this star, in addition to g modes, turned out to be invalid.  HD43317 is thus a SPB star (rather than a hybrid star).  The seismic modelling indicated stellar models with $M_{\star}{=} 5.8^{+0.1}_{-0.2}$\,$\mathrm{M_{\odot}}$, $X_c{=}0.54^{+0.01}_{-0.02}$, and $f_{\rm ov}{=}0.004^{+0.014}_{-0.002}$ to provide the best description of the observations.  This makes HD\,43317 the most massive g-mode pulsator with successful seismic modelling to date.

Using the model frequencies of the best fitted MESA model, we compared the shift of zonal pulsation mode frequencies by the Coriolis force, using the traditional approximation and adopting the measured rotation period at the surface, with those due to Lorentz force, following the perturbative approach of \citet{2005A+A...444L..29H}, the simulations of \citet{{2008MNRAS.386.1947B}}, and our new estimate of the surface field value of $B_{\rm dip} = 1312 \pm 332$\,G.  The maximal magnetic frequency shift was almost always an order of magnitude smaller than the shift caused by the Coriolis force due to a uniform rotation with a period of $0.897673$\,d. Hence, under the adopted approximations, magnetism was a secondary effect compared to rotation when computing pulsation mode frequencies of HD\,43317.  This a posteriori check re-inforced the validity of our modelling approach.

New formalisms for the perturbation of gravito-inertial waves computed with the traditional approximation by a  magnetic field or for the simultaneous non-perturbative treatment of the Coriolis acceleration and of the Lorentz force \citep[e.g.,][]{2011A+A...526A..65M} would be of great value for future seismic modelling of rapidly rotating magnetic massive stars. HD\,43317 can serve as an important benchmark for such future improvement of stellar evolution and pulsation codes, as it is currently the only known magnetic hot star with a relatively rich g-mode frequency spectrum.  Future year-long data sets to be assembled by the NASA TESS mission in its Continuous Viewing zone \citep{2016SPIE.9904E..2BR} and by the ESA PLATO mission \citep{2014ExA....38..249R} will certainly bring more magnetic hot pulsators suitable for asteroseismology.

\begin{acknowledgements}
We are very grateful to the MESA/GYRE development team led by Bill Paxton and Richard Townsend, for their great work on the stellar evolution code MESA and the stellar pulsation code GYRE and for offering these codes in open source to the community.  This work has made use of the SIMBAD database operated at CDS, Strasbourg (France), and of NASA's Astrophysics Data System (ADS).  Part of the research leading to these results has received funding from the European Research Council (ERC) under the European Union’s Horizon 2020 research and innovation programme (grant agreements N$^\circ$670519: MAMSIE and N$^\circ$647383 SPIRE) and the Research Foundation Flanders (FWO, Belgium, under grant agreement G.0B69.13).  The computational resources and services used in this work were provided by the VSC (Flemish Supercomputer Center), funded by the Research Foundation - Flanders (FWO) and the Flemish Government - department EWI.

\end{acknowledgements}
\bibliographystyle{aa}
\bibliography{PhD_ADS}

\appendix

\section{MESA inlist files}
\label{sec:appendix_MESA}
The details related to the micro- and macro-physics in MESA are set by means of inlist files, in which the user specifies which parameters are changed from their respective default settings. Below, we provide the two inlist files that were used in this work, namely the inlist\_base and the inlist\_zoom files.  The former accounts for the basic parameter setup and the latter describes how the meshing of the cells in the 1D stellar models has to be performed  Parameters that were not given a value in this list were let to vary during the analysis. \\

\noindent The MESA inlist\_base setup is:\\

\setlength{\leftskip}{.1cm}
\textsf{\&star\_job}\\

\setlength{\leftskip}{0.5cm}
\indent \textsf{show\_log\_description\_at\_start = .false.}\\
\indent \textsf{show\_net\_species\_info = .false.}\\

\indent \textsf{create\_pre\_main\_sequence\_model = .false.}\\
\indent \textsf{pgstar\_flag = .true.}\\

\indent \textsf{write\_profile\_when\_terminate = .false.}\\
\indent \textsf{filename\_for\_profile\_when\_terminate = 'last.prof'}\\

\indent \textsf{history\_columns\_file = 'hist.list'}\\
\indent \textsf{profile\_columns\_file = 'prof.list'}\\

\indent \textsf{change\_lnPgas\_flag = .true.}\\
\indent \textsf{change\_initial\_lnPgas\_flag = .true.}\\
\indent \textsf{new\_lnPgas\_flag = .true.}\\

\indent \textsf{change\_net = .true.}\\
\indent \textsf{new\_net\_name = 'pp\_cno\_extras\_o18\_ne22.net'}\\
\indent \textsf{change\_initial\_net = .true.}\\
\indent \textsf{auto\_extend\_net = .true.}\\

\indent \textsf{initial\_zfracs = 6}\\

\indent \textsf{kappa\_blend\_logT\_upper\_bdy = 4.5d0}\\
\indent \textsf{kappa\_blend\_logT\_lower\_bdy = 4.5d0}\\
\indent \textsf{kappa\_lowT\_prefix = 'lowT\_fa05\_a09p'}\\

\indent \textsf{kappa\_file\_prefix = 'Mono\_a09\_Fe1.75\_Ni1.75'}\\
\indent \textsf{kappa\_CO\_prefix = 'a09\_co'}\\

\indent \textsf{relax\_Y = .true.}\\
\indent \textsf{change\_Y = .true.}\\
\indent \textsf{relax\_initial\_Y = .true.}\\
\indent \textsf{change\_initial\_Y = .true.}\\
\indent \textsf{new\_Y = }\\

\indent \textsf{relax\_Z = .true.}\\
\indent \textsf{change\_Z = .true.}\\
\indent \textsf{relax\_initial\_Z = .true.}\\
\indent \textsf{change\_initial\_Z = .true.}\\
\indent \textsf{new\_Z = 0.014}\\

\setlength{\leftskip}{.1cm}
\indent \textsf{/ !end of star\_job namelist}\\

\textsf{\&controls}\\

\setlength{\leftskip}{0.5cm}
\indent \textsf{initial\_mass = }\\
\indent \textsf{log\_directory = }\\

\indent \textsf{mixing\_length\_alpha = 2.0}\\

\indent \textsf{set\_min\_D\_mix = .true.}\\
\indent \textsf{min\_D\_mix = }\\

\indent \textsf{overshoot\_f0\_above\_burn\_h\_core = 0.005}\\
\indent \textsf{overshoot\_f\_above\_burn\_h\_core = }\\

\indent \textsf{max\_years\_for\_timestep = 2.0d5}\\
\indent \textsf{varcontrol\_target = 5d-4}\\

\indent \textsf{delta\_lg\_XH\_cntr\_max = -1}\\
\indent \textsf{delta\_lg\_XH\_cntr\_limit = 0.05}\\

\indent \textsf{alpha\_semiconvection = 0.01}\\

\indent \textsf{write\_pulse\_info\_with\_profile = .true.}\\
\indent \textsf{pulse\_info\_format = 'GYRE'}\\

\indent \textsf{xa\_central\_lower\_limit\_species(1) = 'h1'}\\
\indent \textsf{xa\_central\_lower\_limit(1) = 1d-3}\\
\indent \textsf{when\_to\_stop\_rtol = 1d-3}\\
\indent \textsf{when\_to\_stop\_atol = 1d-3}\\

\indent \textsf{terminal\_interval = 25}\\
\indent \textsf{write\_header\_frequency = 1}\\
\indent \textsf{photostep = 500}\\
\indent \textsf{history\_interval = 1}\\
\indent \textsf{write\_profiles\_flag = .false.}\\
\indent \textsf{mixing\_D\_limit\_for\_log = 1d-4}\\

\indent \textsf{use\_Ledoux\_criterion = .true.}\\
\indent \textsf{D\_mix\_ov\_limit = 0d0}\\

\indent \textsf{which\_atm\_option = 'photosphere\_tables'}\\

\indent \textsf{calculate\_Brunt\_N2 = .true.}\\

\indent \textsf{cubic\_interpolation\_in\_Z = .true.}\\
\indent \textsf{use\_Type2\_opacities = .false.}\\
\indent \textsf{kap\_Type2\_full\_off\_X = 1d-6}\\
\indent \textsf{kap\_Type2\_full\_on\_X = 1d-6}\\

\setlength{\leftskip}{.1cm}
\indent \textsf{/ ! end of controls namelist}\\

\setlength{\leftskip}{0cm}

\noindent The MESA inlist\_zoom setup is:\\

\setlength{\leftskip}{.1cm}
\textsf{\&star\_job}\\

\setlength{\leftskip}{.1cm}
\indent \textsf{/ !end of star\_job namelist}\\

\textsf{\&controls}\\

\setlength{\leftskip}{0.5cm}

\indent \textsf{mesh\_delta\_coeff = 0.3}\\
\indent \textsf{max\_allowed\_nz = 80000}\\
\indent \textsf{okay\_to\_remesh = .true.}\\
\indent \textsf{max\_dq = 1d-3}\\

\indent \textsf{T\_function2\_weight = 100}\\
\indent \textsf{T\_function2\_param = 2d5}\\
\indent \textsf{log\_kap\_function\_weight = 100}\\

\indent \textsf{R\_function\_weight = 10}\\
\indent \textsf{R\_function\_param = 1d-4}\\

\indent \textsf{R\_function2\_weight = 10}\\
\indent \textsf{R\_function2\_param1 = 1000}\\
\indent \textsf{R\_function2\_param2 = 0}\\

\indent \textsf{xtra\_coef\_above\_xtrans = 0.1}\\
\indent \textsf{xtra\_coef\_below\_xtrans = 0.1}\\
\indent \textsf{xtra\_dist\_above\_xtrans = 0.5}\\
\indent \textsf{xtra\_dist\_below\_xtrans = 0.5}\\

\indent \textsf{mesh\_logX\_species(1) = 'h1'}\\
\indent \textsf{mesh\_logX\_min\_for\_extra(1) = -12}\\
\indent \textsf{mesh\_dlogX\_dlogP\_extra(1) = 0.1}\\
\indent \textsf{mesh\_dlogX\_dlogP\_full\_on(1) = 1d-6}\\
\indent \textsf{mesh\_dlogX\_dlogP\_full\_off(1) = 1d-12}\\

\indent \textsf{mesh\_logX\_species(2) = 'he4'}\\
\indent \textsf{mesh\_logX\_min\_for\_extra(2) = -12}\\
\indent \textsf{mesh\_dlogX\_dlogP\_extra(2) = 0.1}\\
\indent \textsf{mesh\_dlogX\_dlogP\_full\_on(2) = 1d-6}\\
\indent \textsf{mesh\_dlogX\_dlogP\_full\_off(2) = 1d-12}\\

\indent \textsf{xtra\_coef\_czb\_full\_on = 0.9d0}\\
\indent \textsf{xtra\_coef\_czb\_full\_off = 1d0}\\

\indent \textsf{xtra\_coef\_a\_l\_nb\_czb = 0.1}\\
\indent \textsf{xtra\_coef\_a\_l\_hb\_czb = 0.1}\\
\indent \textsf{xtra\_coef\_a\_l\_heb\_czb = 0.1}\\
\indent \textsf{xtra\_coef\_a\_l\_zb\_czb = 0.1}\\

\indent \textsf{xtra\_coef\_b\_l\_nb\_czb = 0.1}\\
\indent \textsf{xtra\_coef\_b\_l\_hb\_czb = 0.01}\\
\indent \textsf{xtra\_coef\_b\_l\_heb\_czb = 0.1}\\
\indent \textsf{xtra\_coef\_b\_l\_zb\_czb = 0.1}\\

\indent \textsf{xtra\_coef\_a\_u\_nb\_czb = 0.1}\\
\indent \textsf{xtra\_coef\_a\_u\_hb\_czb = 0.1}\\
\indent \textsf{xtra\_coef\_a\_u\_heb\_czb = 0.1}\\
\indent \textsf{xtra\_coef\_a\_u\_zb\_czb = 0.1}\\

\indent \textsf{xtra\_coef\_b\_u\_nb\_czb = 0.1}\\
\indent \textsf{xtra\_coef\_b\_u\_hb\_czb = 0.1}\\
\indent \textsf{xtra\_coef\_b\_u\_heb\_czb = 0.1}\\
\indent \textsf{xtra\_coef\_b\_u\_zb\_czb = 0.1}\\

\indent \textsf{xtra\_dist\_a\_l\_nb\_czb = 0.5}\\
\indent \textsf{xtra\_dist\_a\_l\_hb\_czb = 0.5}\\
\indent \textsf{xtra\_dist\_a\_l\_heb\_czb = 0.5}\\
\indent \textsf{xtra\_dist\_a\_l\_zb\_czb = 0.5}\\

\indent \textsf{xtra\_dist\_b\_l\_nb\_czb = 0.5}\\
\indent \textsf{xtra\_dist\_b\_l\_hb\_czb = 0.01}\\
\indent \textsf{xtra\_dist\_b\_l\_heb\_czb = 0.5}\\
\indent \textsf{xtra\_dist\_b\_l\_zb\_czb = 0.5}\\

\indent \textsf{xtra\_dist\_a\_u\_nb\_czb = 0.5}\\
\indent \textsf{xtra\_dist\_a\_u\_hb\_czb = 0.5}\\
\indent \textsf{xtra\_dist\_a\_u\_heb\_czb = 0.5}\\
\indent \textsf{xtra\_dist\_a\_u\_zb\_czb = 0.5}\\

\indent \textsf{xtra\_dist\_b\_u\_nb\_czb = 0.5}\\
\indent \textsf{xtra\_dist\_b\_u\_hb\_czb = 0.5}\\
\indent \textsf{xtra\_dist\_b\_u\_heb\_czb = 0.5}\\
\indent \textsf{xtra\_dist\_b\_u\_zb\_czb = 0.5}\\

\indent \textsf{xtra\_coef\_os\_full\_on = 1d0}\\
\indent \textsf{xtra\_coef\_os\_full\_off = 1d0}\\

\indent \textsf{xtra\_coef\_os\_above\_nonburn = 0.1}\\
\indent \textsf{xtra\_coef\_os\_below\_nonburn = 0.1}\\
\indent \textsf{xtra\_coef\_os\_above\_burn\_h = 0.1}\\
\indent \textsf{xtra\_coef\_os\_below\_burn\_h = 0.1}\\
\indent \textsf{xtra\_coef\_os\_above\_burn\_he = 0.1}\\
\indent \textsf{xtra\_coef\_os\_below\_burn\_he = 0.1}\\
\indent \textsf{xtra\_coef\_os\_above\_burn\_z = 0.1}\\
\indent \textsf{xtra\_coef\_os\_below\_burn\_z = 0.1}\\

\indent \textsf{xtra\_dist\_os\_above\_nonburn = 0.5}\\
\indent \textsf{xtra\_dist\_os\_below\_nonburn = 0.5}\\
\indent \textsf{xtra\_dist\_os\_above\_burn\_h = 0.5}\\
\indent \textsf{xtra\_dist\_os\_below\_burn\_h = 0.5}\\
\indent \textsf{xtra\_dist\_os\_above\_burn\_he = 0.5}\\
\indent \textsf{xtra\_dist\_os\_below\_burn\_he = 0.5}\\
\indent \textsf{xtra\_dist\_os\_above\_burn\_z = 0.5}\\
\indent \textsf{xtra\_dist\_os\_below\_burn\_z = 0.5}\\

\indent \textsf{convective\_bdy\_weight = 1d1}\\
\indent \textsf{convective\_bdy\_dq\_limit = 1d-3}\\
\indent \textsf{convective\_bdy\_min\_dt\_yrs = 1d-3}\\

\indent \textsf{remesh\_dt\_limit = -1}\\

\indent \textsf{remesh\_log\_L\_nuc\_burn\_min = -50}\\

\indent \textsf{num\_cells\_for\_smooth\_brunt\_B = 0}\\
\indent \textsf{num\_cells\_for\_smooth\_gradL\_composition\_term = 0}\\

\indent \textsf{P\_function\_weight = 30}\\
\indent \textsf{T\_function1\_weight = 75}\\

\indent \textsf{xa\_function\_species(1) = 'h1'}\\
\indent \textsf{xa\_function\_weight(1) = 80}\\
\indent \textsf{xa\_function\_param(1) = 1d-2}\\

\indent \textsf{xa\_function\_species(2) = 'he4'}\\
\indent \textsf{xa\_function\_weight(2) = 80}\\
\indent \textsf{xa\_function\_param(2) = 1d-2}\\

\setlength{\leftskip}{.1cm}
\indent \textsf{/ ! end of controls namelist}\\

\setlength{\leftskip}{0cm}

\section{GYRE inlist file}
\label{sec:appendix_GYRE}

The computations for the theoretical frequencies and their relevant parameters by GYRE are also governed by input files.  We provide the settings contained in the input files for this work below.  Again, parameters that were varied in the analysis do not have a value assigned to them in this list. \\

\textsf{\&constants}\\
\setlength{\leftskip}{.1cm}
\indent \textsf{/}\\

\textsf{\&model}\\
\setlength{\leftskip}{0.5cm}
\indent \indent \textsf{model\_type = 'EVOL'}\\
\indent \indent \textsf{file = ' '}\\
\indent \indent \textsf{file\_format = 'MESA'}\\
\indent \indent \textsf{reconstruct\_As = .False.}\\
\indent \indent \textsf{uniform\_rotation= .True.}\\
\indent \indent \textsf{Omega\_uni = }\\
\setlength{\leftskip}{.1cm}
\indent \textsf{/}\\

\textsf{\&osc}\\
\setlength{\leftskip}{0.5cm}
\indent \indent \textsf{outer\_bound = 'ZERO'}\\
\indent \indent \textsf{rotation\_method = 'TRAD'}\\
\setlength{\leftskip}{.1cm}
\indent \textsf{/}\\

\textsf{\&mode}\\
\setlength{\leftskip}{0.5cm}
\indent \indent \textsf{l = }\\
\indent \indent \textsf{m = }\\
\indent \indent \textsf{n\_pg\_min = -75}\\
\indent \indent \textsf{n\_pg\_max = -1}\\
\setlength{\leftskip}{.1cm}
\indent \textsf{/}\\

\textsf{\&num}\\
\setlength{\leftskip}{0.5cm}
\indent \indent \textsf{ivp\_solver = 'MAGNUS\_GL4'}\\
\setlength{\leftskip}{.1cm}
\indent \textsf{/}\\

\textsf{\&scan}\\
\setlength{\leftskip}{0.5cm}
\indent \indent \textsf{grid\_type = 'INVERSE'}\\
\indent \indent \textsf{grid\_frame = 'COROT\_I'}\\
\indent \indent \textsf{freq\_units = 'PER\_DAY'}\\
\indent \indent \textsf{freq\_frame = 'INERTIAL'}\\
\indent \indent \textsf{freq\_min = }\\
\indent \indent \textsf{freq\_max = }\\
\indent \indent \textsf{n\_freq = 400}\\
\setlength{\leftskip}{.1cm}
\indent \textsf{/}\\

\textsf{\&shoot\_grid}\\
\setlength{\leftskip}{0.5cm}
\indent \indent \textsf{op\_type = 'CREATE\_CLONE'}\\
\setlength{\leftskip}{.1cm}
\indent \textsf{/}\\

\textsf{\&recon\_grid}\\
\setlength{\leftskip}{0.5cm}
\indent \indent \textsf{op\_type = 'CREATE\_CLONE'}\\
\setlength{\leftskip}{.1cm}
\indent \textsf{/}\\

\textsf{\&shoot\_grid}\\
\setlength{\leftskip}{0.5cm}
\indent \indent \textsf{op\_type = 'RESAMP\_CENTER'}\\
\indent \indent \textsf{n = 12}\\
\setlength{\leftskip}{.1cm}
\indent \textsf{/}\\

\textsf{\&shoot\_grid}\\
\setlength{\leftskip}{0.5cm}
\indent \indent \textsf{op\_type = 'RESAMP\_DISPERSION'}\\
\indent \indent \textsf{alpha\_osc = 5}\\
\indent \indent \textsf{alpha\_exp = 1}\\
\setlength{\leftskip}{.1cm}
\indent \textsf{/}\\

\textsf{\&recon\_grid}\\
\setlength{\leftskip}{0.5cm}
\indent \indent \textsf{op\_type = 'RESAMP\_CENTER'}\\
\indent \indent \textsf{n = 12}\\
\setlength{\leftskip}{.1cm}
\indent \textsf{/}\\

\textsf{\&recon\_grid}\\
\setlength{\leftskip}{0.5cm}
\indent \indent \textsf{op\_type = 'RESAMP\_DISPERSION'}\\
\indent \indent \textsf{alpha\_osc = 5}\\
\indent \indent \textsf{alpha\_exp = 1}\\
\setlength{\leftskip}{.1cm}
\indent \textsf{/}\\

\textsf{\&output}\\
\setlength{\leftskip}{0.5cm}
\indent \indent \textsf{summary\_file = ' '}\\
\indent \indent \textsf{summary\_file\_format = 'TXT'}\\
\indent \indent \textsf{summary\_item\_list = 'M\_star, R\_star, beta, l, n\_pg, omega, freq, freq\_units, E\_norm'}\\
\indent \indent \textsf{mode\_prefix = ''}\\
\indent \indent \textsf{mode\_file\_format = 'HDF'}\\
\indent \indent \textsf{mode\_item\_list = 'l, beta, n\_pg, omega, freq, freq\_units, x, xi\_r, xi\_h, K'}\\
\indent \indent \textsf{freq\_units = 'PER\_DAY'}\\
\setlength{\leftskip}{.1cm}
\indent \textsf{/}\\

\end{document}